\documentclass[12pt,preprint]{aastex}
\shortauthors{Carciofi \& Bjorkman}
\shorttitle{NLTE Monte Carlo Radiative Transfer}

\newcommand{\teff}{T_{\rm eff}}
\newcommand{\um}{\;\mu\rm m}
\newcommand{\bw}{\bar{\varpi}}

\begin{document}

\title{NLTE Monte Carlo Radiative Transfer:  I. The Thermal Properties of Keplerian Disks around Classical Be Stars
 }

\author{A. C. Carciofi}
\affil{Instituto de Astronomia, Geof\'{i}sica e Ci\^encias Atmosf\'ericas, Rua do Mat\~ao 1226, Cidade Universit\'aria, 05508-900, S\~ao Paulo, SP, BRAZIL}
\email{carciofi@usp.br} 

\and
\author{J. E. Bjorkman}
\affil{Ritter Observatory, M.S. 113, Dept. of Physics and Astronomy, University of Toledo, Toledo, OH 43606-3390} 

\begin{abstract}

We present a 3-D NLTE Monte Carlo radiative transfer code that we use to study the temperature and ionization structure of Keplerian disks around Classical Be stars. The method we employ is largely similiar to the Monte Carlo transition probability method developed by Lucy.  Here we present a simplification of his method that avoids the use of the macro atom concept.
%However, here we present an alternative to Lucy's method that avoids the use of the macro atom concept.
Our investigations of the temperature structure of Be star disks show that the disk temperature behavior is a hybrid between the behavior of Young Stellar Object (YSO) disks and Hot Star winds. The optically thick inner parts of Be star disks have temperatures that are similar to YSO disks, while the optically thin outer parts are like stellar winds.
Thus, the temperature at the disk midplane initially drops, reaching a minimum at 3--5 stellar radii, after which it rises back to the optically thin radiative equilibrium temperature at large distances. On the other hand, the optically thin upper layers of the disk are approximately isothermal --- a behavior that is analogous to the hot upper layers of YSO disks.  Interestingly, unlike the case of YSO disks, we find that disk flaring has little effect on the temperature structure of Be star disks.
We also find that the disks are fully ionized, as expected, but there is an ionization minimum in the vicinity of the temperature minimum. The deficit of photoionization at this location makes it the most likely site for the low ionization state lines (e.g., \ion{Fe}{2}) that produce the shell features observed in Be stars.  Finally, we find that, despite the complex temperature structure, the infrared excess is 
well-approximated by an equivalent isothermal disk model whose temperature is about 60\% of
the stellar temperature.  
This is largely because, at long wavelengths, the effective photosphere of the disk is located in its isothermal regions.

\end{abstract}

\keywords{radiative transfer --- stars: emission line, Be --- circumstellar matter}

\section{Introduction \label{introduction}}

Circumstellar disks occur in a wide variety of astrophysical systems, ranging from Young Stellar Objects (YSOs) to rapidly rotating main sequence B stars (the Be stars) to highly evolved systems like cataclysmic variables, black holes, and perhaps asymptotic giant branch (ABG) stars and luminous blue variables (LBVs).  In general, such disks tend to form whenever there is an
accreting or outflowing circumstellar envelope in which the radial flow speed 
is small compared to the rotation speed.  These disks reprocess the starlight, 
typically producing a strong infrared (IR) excess.  Furthermore, these disks 
can be quite optically thick at some wavelengths, while being optically thin at others. Consequently, the radiation field at a given point can be quite 
non-local and anisotropic.  In this paper, we develop a non-local thermodynamic equilibrium (NLTE) Monte Carlo radiation transfer code that may be applied to 
study differentially moving three-dimensional circumstellar envelopes.  As a 
particular application of this code, we study the temperature structure of the 
disks around Be stars.

The circumstellar environment of Be stars has been studied extensively in the past few decades \citep[see][for a recent review]{por03}.  Both observational and theoretical evidence strongly suggests that most of the circumstellar matter is contained in a very dense, geometrically thin disk \citep*{qui97,woo97}.  Since the stellar temperature is quite high, 
these disks are expected to be completely ionized and devoid of dust, so the disk opacity (primarily from hydrogen and free electrons) is much better understood than, for example, YSO disks (where the dominant opacity source is dust, whose size distribution, chemical composition, and even spatial distribution is rather uncertain).  What is less well-determined is the detailed structure (density, temperature, ionization fraction, and velocity) of Be star disks.

Kinematic information about the disk can be obtained from the disk emission 
lines.  These line profiles are often double-peaked with full widths (in the 
wings) that exceed the $v \sin i$ of the star by about 20\% \citep{han96}, so 
the disk must rotate faster than the star.  Based on the symmetry of the 
hydrogen Balmer line profiles and, especially the optically thin Fe\,{\sc ii} 
lines, it appears that the expansion velocity, $v_r$, of the disk must be much 
less than its rotation speed, $v_\phi$ \citep{han88,han95,han96,han00}.  
Studies that try to determine the rotation speed as a function of radius are 
less conclusive.  Using a power-law parameterization, $v_\phi \propto r^{-j}$, 
\citet{hum00} find that they cannot distinguish between angular 
momentum-conserving ($j=1$) and Keplerian ($j=1/2$) velocities for optically 
thin lines, but for optically thick lines they find $v_\phi \propto r^{-0.65}$, favoring Keplerian orbital motion.  Theoretical evidence for Keplerian rotation is provided by the observation of periodic variations in the shape of the 
double-peaked emission line profiles.  During these variations, the ampiltudes of the violet and red emission peaks change in an antisymmetric fashion 
(so-called $V/R$ variations).  Such behavior can be explained theoretically by a precessing one-armed density wave \citep{oka91}, but on theoretical grounds such density waves require a Keplerian velocity structure \citep{oka91}. 
We conclude, therefore, that the disk has a Keplerian orbital velocity with 
$v_r \ll v_\phi$.

The IR excess of Be stars arises from free-free (f-f) emission in the disk 
\citep*{geh74}.  Since the f-f opacity increases with wavelength, the 
slope of the IR excess can be used to probe the radial density structure of the
circumstellar envelope \citep{wri75,lam84}.  Assuming an isothermal disk with a radial power law for the density, $\rho \propto r^{-n}$, this technique gives 
density exponents in the range $n=2$--3.5 \citep*{wat86,cot87,wat87}.  It 
should be noted however, that this result depends on the radial structure of 
the disk temperature \citep[see][which shows how the slope of the IR
excess depends on the temperature gradient]{cas77}.  

Although quite sophisticated radiative equilibrium models have been developed
that can determine the 2-D temperature structure of YSO disks and envelopes
\citep[e.g.,][]{dal98,woo02,dul02,whi03,wal04}, we are only now beginning to 
understand the temperature structure of Be star disks.  Models of spherically 
expanding stellar winds suggested that, as a consequence of the Cayrel effect
\citep{cay63}, the wind ionization fractions would ``freeze'' (i.e., be 
constant) at large radii and that the temperature would become isothermal with 
a temperature of about $0.8\teff$ \citep[][however later work by
\citealt{dre89}, who included cooling by heavy elements, shows the temperature is not particularly isothermal, slowly falling to less than 60\% $\teff$ at large distances]{kle78}.  In any case, the usual assumption for the disk
temperature was to hope for a similar situation and to assume that the disk was isothermal with a temperature $T_d=0.8\teff$ \citep[e.g.,][]{wat86}.  The first attempts to calculate the disk temperatures were performed by \citet{mil98}, who implemented the radiative equilibrium condition within the \citet{poe78} code.  Note however, that the \citeauthor{poe78} code does not perform a full 
two-dimensional solution to the radiative transfer problem; it instead employs several approximations to estimate the hydrogen level populations.  In
subsequent papers, \citet{mil99a} approximated the effects of diffuse emission by employing the ``on-the-spot approximation'', while \citet*{jon04} included cooling by Fe line emission.  In general, they find that the disk temperature in the mid-plane is quite low,  while the upper layers of the disk are hot.  This fundamental work shows that the disk is certainly not isothermal, bringing into question the \citeauthor{wat87} density results.   For this reason, \citet{mil99b} calculated the temperature for the \citet{wat86} density model. They concluded that the average disk temperature is a factor of 2--3 lower than that assumed by \citeauthor{wat86}.  An additional effect \citep[not addressed by][]{mil99b} is that the non-isothermal temperature structure may alter
the slope of the IR continuum, thereby changing the estimates of the radial
density exponent, $n$.

In this paper, we build upon and extend the results of 
\citet{mil98,mil99a,mil99b} to study more fully the temperature and ionization 
structure of the disk.  In particular, we investigate what physically 
determines the temperature behavior of the disk, and we compare this behavior 
to that of YSO accretion disks and stellar winds.  We also examine the effects
of a non-isothermal disk temperature on the IR spectral energy distribution 
(SED) of Be stars.  To do so, we have developed a Monte Carlo radiative 
transfer code (presented in section~\ref{MC}) that fully solves the 3-D NLTE 
problem and radiative equilibrium temperature.  In section~\ref{tests}, we 
present test calculations used to verify both the method and the computer code. The disk model used in this paper is described in Section~\ref{diskmodel}. In 
Section~\ref{results}, we investigate the dependence of the temperature, 
hydrogen level populations, and ionization fractions on model parameters such 
as the disk density and geometry. Similarly, we investigate how the temperature structure alters the slope of the IR excess.  Finally, we summarize our results in Section~\ref{discussion}.

%-----------------------
\section{The Monte Carlo Code \label{MC}}

One of the greatest assets of the Monte Carlo (MC) method is its flexibility 
and ease of implementation, especially as the complexity of the system 
increases. Although traditional radiative transfer methods are much more 
computationally efficient for simple configurations (e.g. plane-parallel or 
spherical geometries), for more complex situations (e.g., 3-D geometries, 
anisotropic scattering, and complex frequency redistribution processes) the MC 
method is frequently the only viable alternative.  Furthermore, the relative 
efficiency of Monte Carlo methods increases as the number of spatial 
dimensions, frequencies, and anisotropy of the radiation field increases.

\begin{figure}[bp]
\plotone{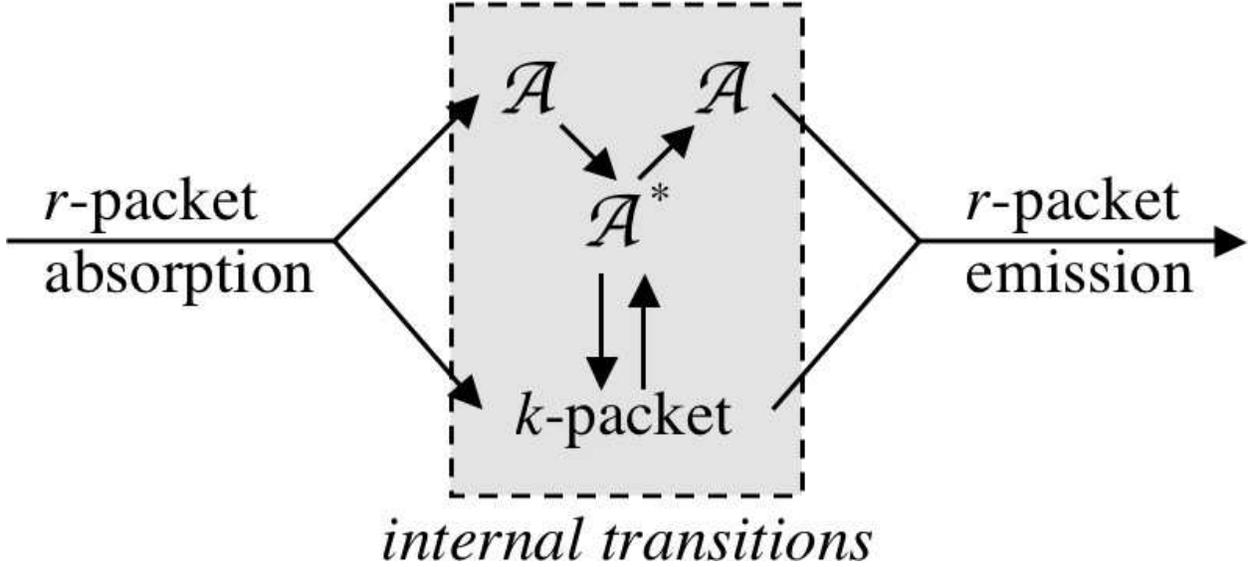}
\figcaption[]{Photon-Macro Atom Interaction.  In Lucy's method \citep{luc02} when a photon ($r$-packet) is absorbed, it either activates the macro atom ($A \rightarrow A^*$), or it is converted into a $k$-packet (free electron energy). If it activates the macro atom, the macro atom undergoes internal transitions (possibly producing a $k$-packet and deactivating) until a photon packet is eventually reemitted ($A^* \rightarrow A$).  If a $k$-packet is produced, the $k$-packet can either reactivate the macro atom, or it can emit a photon packet directly.  In any case, the absorption of a photon eventually leads to the reemission of a new photon.
\label{Diagram}}
\end{figure}

Recently, \citet[][hereafter L02, L03, and L05]{luc02,luc03,luc05} developed a Monte
Carlo transition probability method to solve the statistical equilibrium 
problem.  His method employs indivisible energy packets to describe the 
exchange of energy between photons, \emph{macro atoms} (an ensemble 
representation of the atomic excitation throughout a finite volume element), 
and free electrons.  The energy packets contain either radiative energy ($r$-
packets) or kinetic energy ($k$-packets).  When an energy packet (either
radiative or kinetic) is absorbed by a macro atom, the macro atom randomly
performs internal transitions (using an effective transition probability 
array) between its energy states until it emits a new energy packet (either 
kinetic or radiative); see Figure~\ref{Diagram}.  Since any photon absorption 
is eventually followed by emission of an equal energy photon packet, this
method automatically enforces radiative equilibrium (in the rest frame 
of the gas).  The key ingredient of this method is the effective transition 
probability array, and L02 shows how to choose these probabilities so that the 
internal transitions of the macro atoms will reproduce the statistical 
equilibrium equations.  Since the transition probabilities (effectively they 
are the net rates between levels) depend on both the values of the mean 
intensity in each line and the level populations, their solution requires 
iteration.  

L02 shows that this technique correctly reproduces the line and continuum 
emissivities, the ``$\Lambda$-iteration'' converges, and on average there 
are only about two internal transitions per photon-macro atom interaction.  
In subsequent papers (L03 and L05), Lucy applied this method to the non-trivial 
problem of calculating the hydrogen level populations and electron 
temperature in the expanding shell of a type II supernova.  He also studied 
the convergence behavior and demonstrated that the correct solution can be 
achieved with a reasonable amount of computing time.

In 2002-2004, we independently developed a method for solving both the 
statistical and radiative equilibrium in circumstellar winds and disks of hot 
stars. Our method is largely similar to the one described in L03, in that we 
measure the rates (radiative, heating, and cooling) using essentially identical methods --- path length sampling, \citep[see][hereafter L99]{luc99}, so rather 
than reproducing those formulae here, we simply refer the reader to sections 5 
and 6 of L03 and section 2 of L05.
%probably because we based our method on ideas presented in L99.
However our method differs from Lucy's in one key area: the photon-atom
interactions.
%, which has the potential of making our method easier to implement. 
In the following, we present a short description of our method, focusing on the differences with Lucy's method.

The essence of any method for solving the radiative transfer problem is to
correctly describe the radiation field everywhere in the so-called interaction region (the physical medium wherein the radiation propagates). In the MC 
method, this is accomplished statistically by simulating the random propagation of $N$ photon packets (PPs) through the medium, where they are scattered, 
absorbed, and reemitted until they eventually leave the interaction region. A 
convenient choice is to emit equal-energy monochromatic PPs from the star (as a consequence, packets of differing frequencies contain different numbers of 
physical photons).  The condition of radiative equilibrium then implies that whenever a PP is absorbed, it must be reemitted at the same location with the same energy (in the rest frame of the fluid; in the stellar frame, the packet energy changes, owing to the doppler shift between the different reference frames).  This device automatically ensures flux conservation; i.e., the flux 
is everywhere divergence free \citep{luc99,bjo01}.  Although the (rest
frame) packet energy does not change, the new direction and frequency of the 
PP must be chosen to reproduce the emissivity $j_\nu$ of the medium at that 
location. 
 
To solve the statistical equilibrium rate equations, the following procedure 
is employed.  Initially, the interaction region is divided into a number of cells.  In each cell, the gas state variables (density, velocity gradient,
kinetic temperature, and level populations) are assumed to be constant.
The simulation starts with an initial estimate of the state variables for each cell.  While the simulation is performed, the radiative rates, as well as the heating rate, of the gas are sampled as photons pass through each cell. At the end of the MC simulation, the accumulated radiative and heating rates are used to update the cell level populations and kinetic temperature.  A new 
simulation then is performed using the updated values of the state variables. This procedure is repeated until the state variables converge to their equilibrium values for all cells.

From the above description, it is evident that two key points are necessary for the method to succeed. First, there must be an efficient way to sample all 
the radiative transitions and heating rates.  An elegant and efficient solution for this was proposed by L99 and further developed in L03. In section \ref{bb_sample}, we describe how to more efficiently sample the bound-bound (b-b) rates.  The second critical point is that the MC simulated 
radiation field has to statistically reproduce the ``true'' radiation field of the system. Because every PP absorption is immediately followed by \emph{in situ} reemission, this condition reduces to sampling correctly the frequency and direction of the reemitted PPs.

Lucy's method (fig.~\ref{Diagram}) simulates the interaction of the radiation with matter by tracking transitions of his macro-atoms.  When a PP ($r$-packet) undergoes an absorption event (determined by the opacities), it will either activate a macro atom (for instance, through b-b or bound-free transitions) or be converted into a $k$-packet (for example, via f-f absorption).  A series of internal transitions (controlled by the MC transition probabilities, L02) eventually leads to the conversion of the energy back into an $r$-packet.  Thus, Lucy's method is a scheme for sampling the frequency of the reemitted PPs.

The approach we adopt in this paper differs from Lucy's in that we do not use his MC transition probabilities to sample the frequency of the reemitted PP. Instead, we bypass his macro-atoms and directly sample the frequency from the total emissivity of all possible emission processes.  Thus our emission process
is completely unlinked from the absorption process.  For example, a photon may
be absorbed by a b-b transition and then reemitted by a f-f
process.  This may seem peculiar, but all that matters is that at the
end of the simulation, we produce the correct total luminosity and frequency
distribution for each process.

For a given cell $j$, the total luminosity of a given process $k$ is
\begin{equation}
L^k_j = \int_{V_j} dV \int_0^\infty 4\pi j^k_\nu\,d\nu. \label{process_emissivity}
\end{equation}
where $j^k_\nu$ is  the emissivity of process $k$ and the first integral is performed over  the cell volume $V_j$. 
%In Appendix~\ref{App_Emissivities} we write down the expressions for the emissivities of all processes considered (professor, probably this appendix in unnecessary because the expressions for the emissivity are well known).
When a PP is absorbed in cell $j$, we select the physical process $k$ that  reemits the PP, determined by the minimum value of $k$ satisfying 
\begin{equation}
\xi < {\sum_{i=1}^{k} L^i_j}/{L_j}, \label{sample_process}
\end{equation}
where $\xi$ is a uniform random number in the interval $(0,1)$, and the total luminosity of the cell 
\begin{equation}
L_j \equiv \sum_i L^i_j. \label{cell_emissivity}
\end{equation}
Once the physical process is chosen, the frequency of the emitted PP is sampled using standard MC techniques (see section 4 of L03).

From the point of view of computer coding, our method has the advantage of
being easier to implement (but one should note that the implementation 
of macro-atoms does not require a particularly large coding effort).
Since we dispense with the macro-atom device, we do not need to track their internal transitions, which in principle increases the relative speed of our code (but, this may not be a large effect since there were only two internal transitions per interaction for the case studied by L02).  Similarly, there is no need to convert energy packets back and forth between $r$-packets and $k$-packets.
%We point out, however, that the above is true only in the absence of non-radiative heating. If additional energy transport mechanisms are employed, the
On the other hand, it is not clear which method will converge faster.  Both methods use the previous line intensities to calculate their reemission probabilities, and the two methods are rigorously equivalent for a system already in equilibrium.  However, out of equilibrium, the two
methods differ.  Our emissivities are constant during a given iteration,
but Lucy's method uses conditional transition probabilities, which may 
somewhat adapt the emissivities to the new values of the line intensities.  We 
suspect, therefore, that Lucy's method might have a somewhat better convergence rate.

Finally, we point out that even using the efficient intensity estimators of 
L03 and L05, the computational demand for 2-D and 3-D problems is enormous, requiring
a parallel-computing cluster to make the run times practical.  Fortunately, the structure of the algorithm is well-suited for parallel computation.  The
problem is ``embarassingly parallel'' --- one simply divides the total number
of photon packets into batches that run independently on each processor. 
For the 2-D models shown in Sections~\ref{results}, we used a 96-node Beowulf cluster (with 1.7 GHz Athlon CPUs).  One iteration for the level populations 
and kinetic temperature requires about 15 minutes on this cluster (using $N = 3 \times 10^7$ PPs). Since the number of iterations required is about 10--15, one model takes a few hours to complete.

%-----------------------
\subsection{Simulations for a Small Frequency Range}

A common problem for full-spectrum MC codes is that the radiation field is typically under-sampled at frequencies where the mean intensity is low. This happens because the number of PPs per frequency interval is proportional to the mean intensity. As a result, calculating the emergent flux at those frequencies requires a very large number of PPs for the simulation.  Similarly, calculating the emergent flux for a very narrow frequency range (as is the case for line profiles) can be problematic because the number of PPs in that range is small.

A related issue is determining the b-b rates required to solve the statistical equilibrium equations, because sampling each transition requires a large number of PPs in a small frequency range. For instance, suppose that the mean intensity at a given point of the interaction region is given by a Planck function. The ratio of resonant H$\alpha$ PPs to the total number of PPs is approximately
\begin{equation}
\frac{N_{\rm{H}\alpha}}{N} \approx \frac{\left(v_{\rm max}/c\right) \nu_{\rm{H}\alpha } B_\nu(\nu_{\rm{H}\alpha},T)}{\sigma T^4},
\end{equation}
where $v_{\rm{max}}$ is the maximum macroscopic velocity of the gas and $\sigma$ is the Stephan-Boltzmann constant. For $v_{\rm{max}}=1000 \;{\rm km\,s}^{-1}$ and $T=16000\;\rm K$, this ratio is about $10^{-4}$, which illustrates the low number of PPs available for sampling the b-b rates. If one considers that the main source of exciting PPs comes from the external radiation source (the star), and that those PPs would be split over the very large number of cells necessary for 2-D or 3-D problems, it is easy to see that a huge number of PPs would be necessary to correctly sample the b-b 
rates. The solution to this problem is simple: limit the frequency range of the MC simulation to the vicinity of the line of interest, and sample each b-b transition separately. 

Consider a star surrounded by an envelope that is divided into $M$ cells. 
The total energy emitted by the star per unit time in the frequency range $\nu_0$ and $\nu_1$ is 
\begin{equation}
L_\star = 4\pi R_\star^2 \int_{\nu_0}^{\nu_1} \mathcal{F}^{+}_\nu d\nu,
\end{equation}
where $\mathcal{F}^{+}_\nu$ is the outward flux at the stellar surface. Similarly, the luminosity of process $k$ in cell $j$, $L^k_j$, can be calculated from eq.~(\ref{process_emissivity}), using the appropriate frequency range in the integral. The total energy emitted by the cell, $L_j$, is calculated from eq.~(\ref{cell_emissivity}), and the total energy emitted by the envelope is given by
\begin{equation}
L_{\rm env} = \sum_{j=1}^M L_j.
\end{equation}

Given the energies calculated above, the MC calculation proceeds step-by-step as follows:

\begin{enumerate}
\item A new PP is generated. If $\xi < L_\star / \left(L_\star+L_{\rm env} \right)$ the PP is emitted by the star (go to step 2), and if $\xi > L_\star / \left(L_\star+L_{\rm env} \right)$ the PP is emitted by the envelope (go to step 3).
\item The position on the stellar surface is sampled randomly, the PP frequency
is chosen from the stellar flux, $\mathcal{F}^{+}_\nu$, over the relevant 
frequency range, and the propagation direction is sampled from the limb darkening law. Proceed to step~6.
\item The cell $j$ that emits the PP is sampled using the relation $\xi  < \sum_{i=1}^j L_i/L_{\rm env}$. 
\item The process $k$ that emits the PP in cell $j$ is sampled using eq.~(\ref{sample_process}).
\item The position of the PP inside the cell is chosen randomly, and the PP propagation direction is chosen accordingly to the process of emission.
For example, the direction distribution for line emission is given by the
Sobolev escape probability.
\item The PP is propagated through the medium using standard MC techniques until it leaves the envelope. If the PP is absorbed or hits the star, it is destroyed. Return to step~1 until all $N$ PPs are generated.
\end{enumerate}

\subsubsection{Detailed Spectra}
The above procedure can be used for the calculation of emergent spectra, both where the flux is low (e.g., the IR flux in hot stars) and when the frequency range is small (e.g., a detailed calculation of the shape the balmer jump).  Recall that in the radiative equilibrium scheme, a PP may be absorbed and reemited many times before escaping --- a time-consuming process --- while in the above procedure the PP is simply destroyed.  Thus in addition to enabling the calculation of multi-frequency radiative transfer in arbitrary frequency ranges, the above procedure also has the advantage of performing much quicker spectral syntheses.  Furthermore, it automatically includes anisotropic scattering and polarization effects that are difficult to include in traditional post-processing routines.  

\subsubsection{Sampling Bound-Bound Rates \label{bb_sample}}

Finally, we use the above technique to sample the b-b rates in the following way. We first run a standard radiative equilibrium simulation to calculate all the radiative and heating rates, apart from from the b-b rates. Then a series of simulations are performed to sample individually the rates for each b-b transition.  By choosing a frequency range $\left[\nu_{\rm L} (1-2v_{\rm max}/c), \nu_{\rm L} (1+2v_{\rm max}/c) \right]$, where $\nu_{\rm L}$ is the frequency of the transition (i.e., centered on the line of interest), we 
ensure that we properly account for the coupling to all possible processes, 
including other emission lines.

\section{Numerical Tests \label{tests}}

Here we verify our code results and test the convergence of our algorithm.
We present two test cases: 1) a blackbody cavity, which has a simple analytic solution, and 2) the non-trivial problem of a spherically symmetric expanding 
stellar wind.

\subsection{Blackbody Cavity}

This test consists of a spherical cavity whose boundary emits a blackbody spectrum of a given temperature. The cavity is filled with hydrogen with constant density throughout.  Each PP emitted by the cavity wall undergoes a random series of scatterings and absorptions followed by {\it in situ} reemissions as described in section~\ref{MC}, but eventually the PP is absorbed (and destroyed) by the blackbody cavity wall, whereupon another PP 
is emitted until the total number PPs have been run.

A blackbody cavity is an especially convenient test because the gas should
come to complete thermodynamic equilibrium with the wall.  Thus the state variables of the gas depend solely on the temperature of the cavity's wall, $T_{\rm cav}$, and the level populations and ionization fraction should be given by the Boltzmann-Saha distribution.  Furthermore, the local radiation field should be a Planck function, and all radiative rates should attain their LTE values.  Finally, the kinetic temperature of the electrons should also be $T_{\rm cav}$.

\begin{figure}[bp]
\plottwo{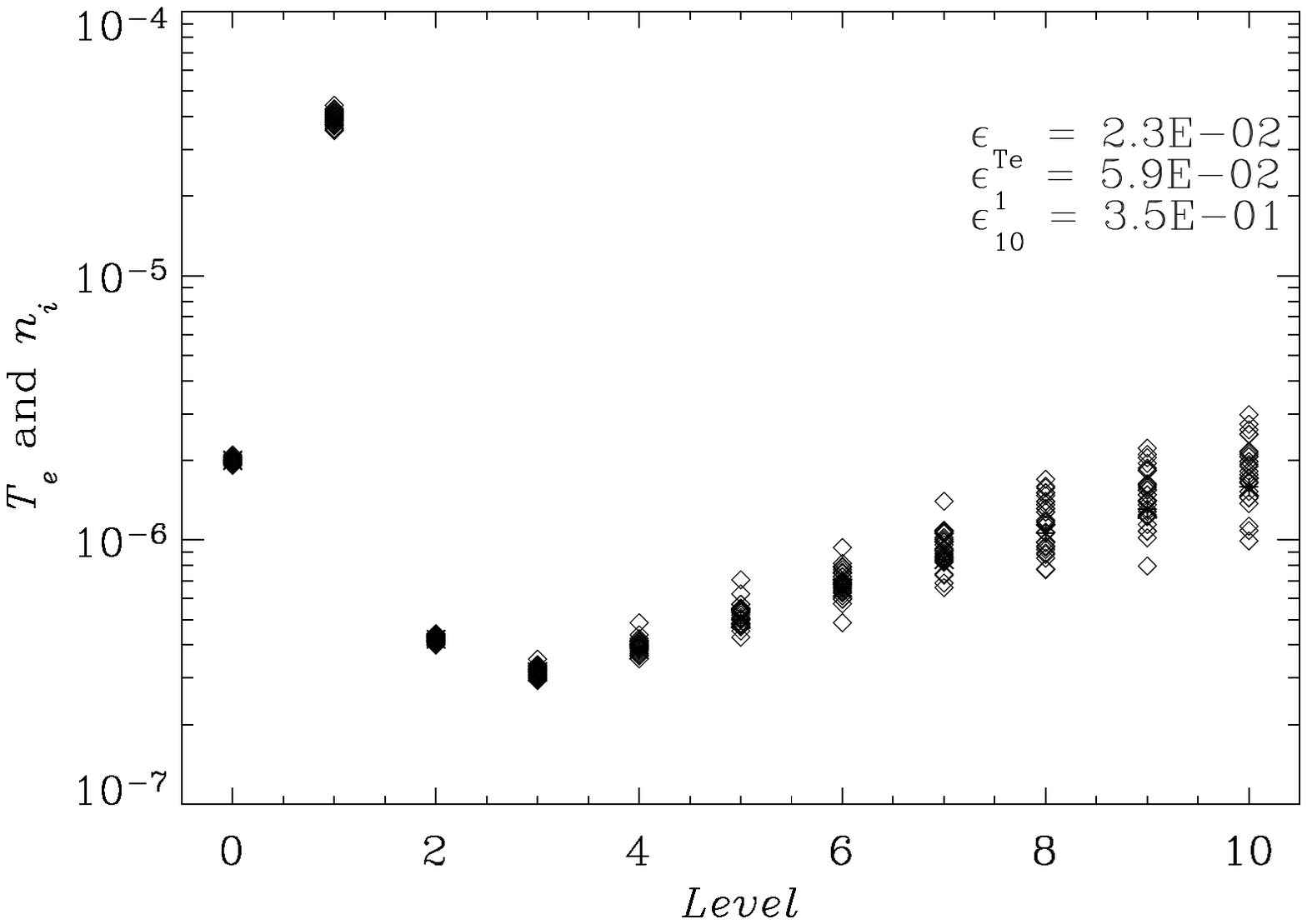}{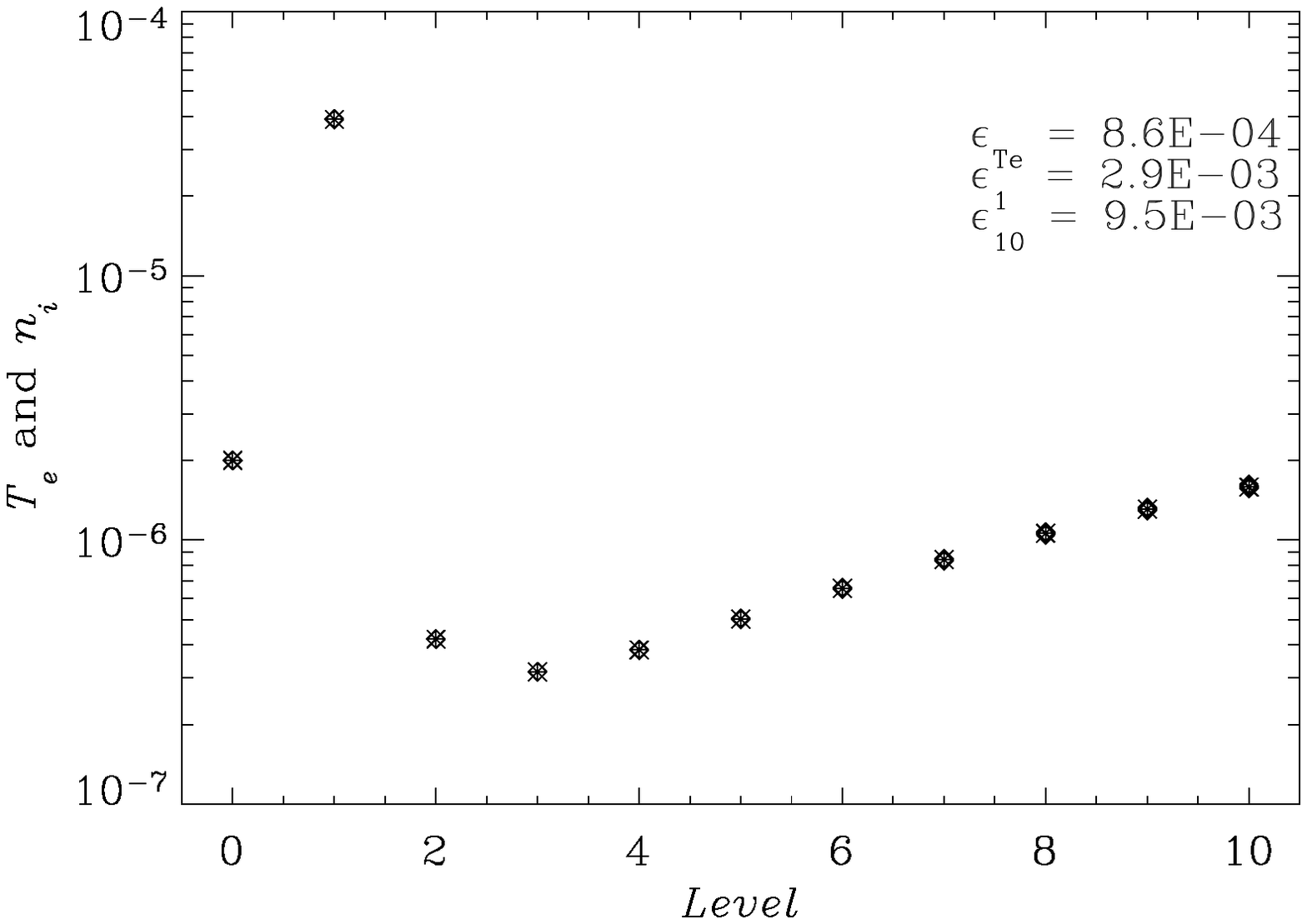}
\figcaption[]{Stability Test. The electron temperature (level 0 corresponds to $T_e/10^{10}$) and hydrogen level populations (up to level 10) are shown for 30 successive iterations.  The left panel shows the results (triangles) for $N=10^4$ PPs, while the right panel is for $N=10^7$ PPs.  Note that the variation of the values is centered on the correct (LTE) values (stars).  The R.M.S. variation, $\epsilon$, for various levels is indicated.
\label{Cavity1}}
\end{figure}

We used this setup to test several aspects of our method. A trivial but important situation is to set the initial level populations and electron temperature to their correct equilibrium values; i.e., $T_e = T_{\rm cav}$ and $n_i = n_i^{\rm *}$, where the asterisk indicates LTE level populations. 
In this case, one expects that, if the rates are being properly sampled, the state variables should oscillate around the LTE values as a result of MC statistical sampling errors.  Furthermore, the magnitude of the oscillation should decrease approximately\footnote{Since there is a nonlinear relationship between the rates and the state variables, the statistical error will not follow exactly a Poisson distribution (unless the fluctuations are small).}  as $1/\sqrt{N}$.  This test is also important because it demonstrates the stability of a converged solution.
The results for 30 iterations with initial conditions $T_{\rm cav} = 20000 \;\rm K$ and hydrogen number density  $n_H = 10^{14} \;\rm cm^{-3}$ are shown in Figure~\ref{Cavity1}.  The simulation in the left panel employed $N = 10^4$ PPs, while the right panel had $N = 10^7$.  As expected, the results for
$T_e$ and $n_i$ oscillate around the correct equilibrium values, and the magnitude of the oscillation decreases as $N$ increases.  The R.M.S. 
relative error
\begin{equation}
\epsilon_S = \left[\frac{1}{n_{\rm it}-1}\sum_{i=1}^{n_{\rm it}} (S_i - S^*)^2/(S^*)^2\right]^{1/2},
\label{error}
\end{equation}
where $S$ is the state variable of interest, is indicated for hydrogen levels 1 and 10, as well as the electron temperature.  Note that the R.M.S. errors decrease roughly by the expected amount ($\sqrt{10^3}$).

\begin{figure}[bp]
\plotone{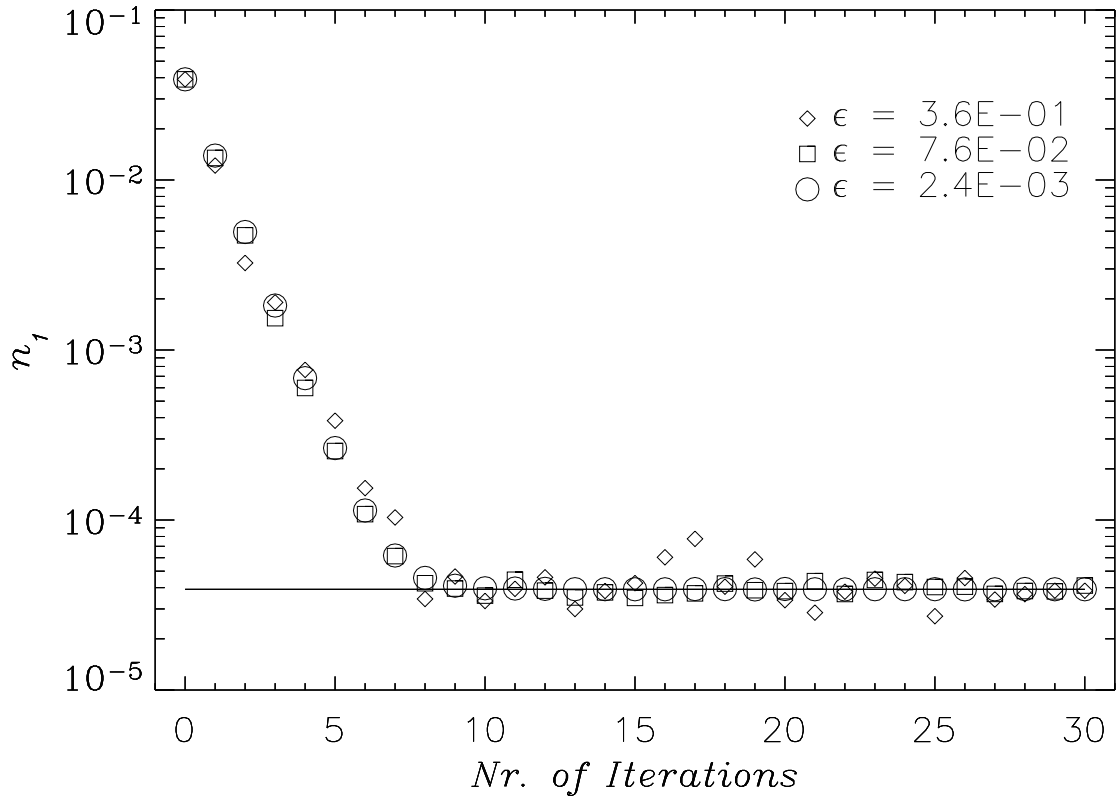}
\figcaption[]{Convergence Test. The hydrogen ground level population is shown for a sequence of 30 iterations (Diamonds: $N=10^3$; Squares: $N=10^4$; Circles: $N=10^7$).  Iteration zero refers to the initial population. The straight line shows the correct level population.
\label{Cavity3}}
\end{figure}

In the next series of tests, we varied the initial values of the state variables to verify that the results did converge to their correct values. As an example, Figure~\ref{Cavity3} shows 30 iterations of the ground level population for three values of $N$ ($10^3$, $10^4$, and $10^7$).  The population was increased initially by a factor of $10^3$, and it converges 
to the correct value (shown as a straight line in the figure) after about 10 iterations.  More importantly, the level population converges even in the presence of the very large sampling errors for the $N=10^3$ case.

The high precision we were able to obtain for the level populations and electron temperature in the blackbody cavity test verifies the validity 
and the correctness of our implementation of the method described in the last section.  In addition, it shows that, as already demonstrated by L02, the random errors intrinsic to the Monte Carlo method do not prevent the convergence of the state variables. It is important to emphasize this last remark because, when the method is applied to 2-D or 3-D geometries, one may easily run into situations for which given parts of the simulation may be under-sampled. An example of such a situation is the equatorial region of optically thick disks, which is shielded from the stellar ionizing radiation by the disk's upper layers.

Finally,  we would like to comment that the blackbody cavity test has proven extremely useful during the development phase of the code.  In particular,
detailed balance occurs in LTE, so one can turn on and off individual processes (e.g., f-f emission and absorption), and the results should not change.  This permits one to test separately each part of the implementation. Similarly, sampling the local radiation field and looking for departures from a Planck function is especially useful for diagnosing errors.

%-----------------------
\subsection{NLTE Test}

\begin{figure}[bp]
\plotone{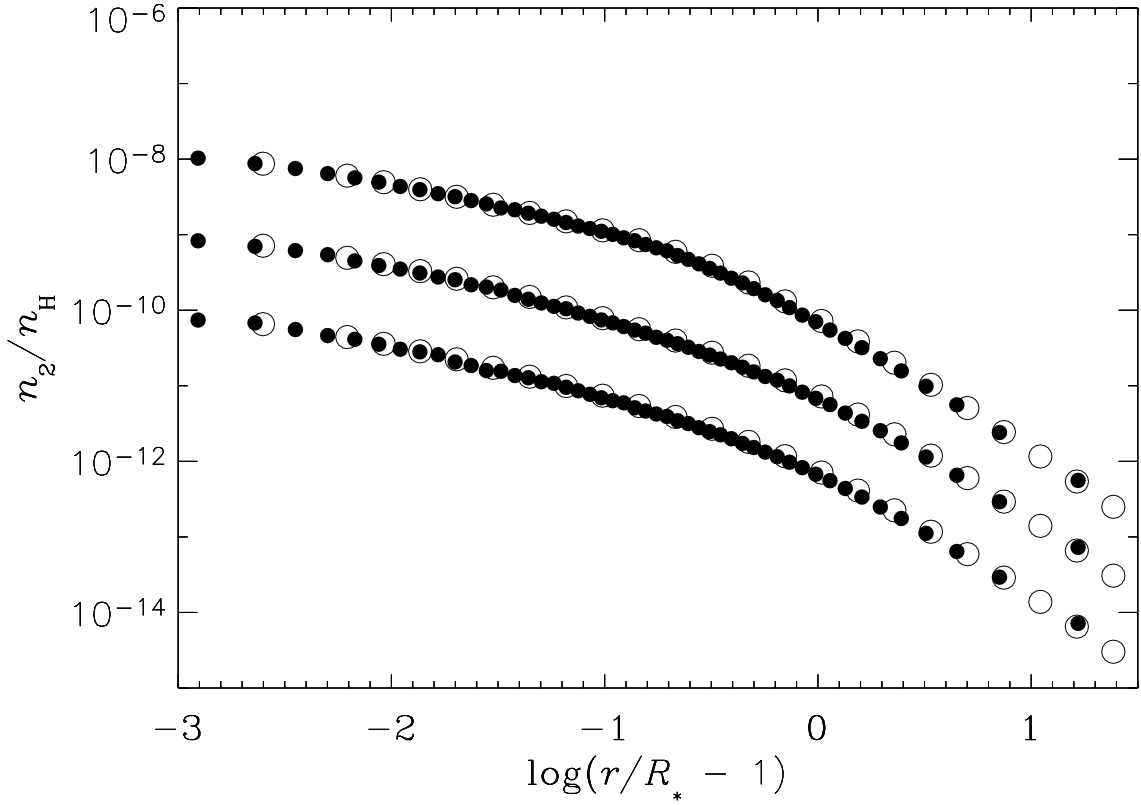}
\figcaption[]{NLTE Validation. Comparison between the results of the  \citet{mac93} code (open circles) and our code (filled circles). Shown are the $n=2$ hydrogen level populations as a function of distance from the star. From bottom to top, the curves are for $\log\dot{M} = -8.7$, $-7.7$ and $-6.7$, respectively.
\label{MF}}
\end{figure}

To assess the NLTE results for our method and to verify the Sobolev escape probabilities for the spectral lines (which modify the rate equations), we applied our code to the 1-D problem of a spherical expanding wind around a hot star, and compared our results to those
given by the well-established code of \citet{mac93}.  We were able to match MacFarlane's results for a wide range of stellar temperatures, wind density, and velocity. An example comparison is given by Figure~\ref{MF}, which shows the $n=2$ hydrogen level populations for a B2\,V star ($\teff = 20000 \rm{K}$,  $R_\star = 6.6 R_\sun$), surrounded by a stellar wind with a $\beta$-law velocity
\begin{equation}
v_r = v_\infty \left(1-\frac{R_\star}{r} \right)^\beta,
\end{equation}
with $v_\infty = \rm 1600\;km\;s^{-1}$ and $\beta=0.8$. 
For all mass-loss rates shown ($\log\dot{M} = -8.7$, $-7.7$ and $-6.7$), the difference between the results is typically less than 0.1\%.

%-----------------------
\section{Disk Model \label{diskmodel}}

As discussed in the introduction, observational evidence indicates that the circumstellar disks of Be stars are geometrically thin, optically thick
disks, whose outflow velocities are small in comparison to their rotation
speeds.  Dynamically this requires that the disk be pressure supported
in the vertical direction and centrifugally supported in the radial direction.  For this reason the disk structure is most likely to be that of a Keplerian hydrostatically supported disk.  How material is injected at Keplerian speed
into this disk is still unclear, but after it is injected, the gas will
spread outward (and inward if it is injected at intermediate radii), owing
to viscous effects \citep{lyn74}.

Determining the self-consistent structure for a Keplerian disk is a 
somewhat complicated problem, because the disk temperature controls its geometry (via the hydrostatic equilibrium equation), which in turn determines the heating and hence temperature of the disk \citep[see][]{ken87}.  Rather than solving the hydrostatic equilibrium equation (which we
plan to investigate in a subsequent paper), we instead assume
that the disk geometry is a  known function.  For our present purposes ---
an initial investigation of the temperature and ionization structure of
the disk --- we feel this is a reasonable approach.

Studies of pre-main-sequence accretion disks \citep[see][for a review]{har98} show that the disk pressure scale height $H = (a/v_\phi)\varpi$ is small (since the orbital speed, $v_\phi$, is typically much greater than the isothermal sound speed $a$), 
so the disk is geometrically thin, and the disk temperature typically falls (roughly) as a power-law with radius.  Assuming the disk is vertically
isothermal, the disk density structure is given, in cylindrical coordinates 
$(\varpi,z,\phi)$, by
\begin{equation}
\rho(\varpi, z) = \rho_0 \left(\frac{R_\star}{\varpi}\right)^n \exp{\left({-\frac{z^2}{2H^2}}\right)} \label{verticalrho}
\end{equation}
\citep[see, for example,][]{bjo97}.  For steady-state isothermal outflow, the
radial density exponent $n=3.5$ \citep{lee91,por99}.  The vertical structure is the familiar Gaussian result for pressure-supported disks with a scaleheight 
\begin{equation}
H(\varpi) = H_0\left( \frac{\varpi}{R_\star}\right)^{\beta}, \label{scaleheight}
\end{equation}
where
\begin{equation}
H_0 =  \frac{a}{v_{\rm crit}}R_\star,
\end{equation}
and $v_{\rm crit}=\sqrt{GM/R_\star}$ is the critical velocity of the star.  
The disk flaring exponent $\beta$ depends on how the disk temperature falls 
with radius.  For optically thick, infinitessimally thin reprocessing disks, 
$T \propto \varpi^{-3/4}$, which gives $\beta=9/8$ \citep[][but see \citealt{dal98}]{ken87}, while for optically thin stellar winds, the temperature is roughly isothermal \citep{kle78}, so $\beta = 1.5$.

The disk flaring parameter $\beta$ controls the shape of the disk surface
and hence its heating versus radius.  As such, it is the primary parameter, along with density, that controls the temperature and ionization structure of the disk.  Consequently for this paper, we adopt the above structure for the circumstellar disk, and let $\beta$ and $\rho_0$ be free parameters.  Some
models of Be star disks assume the disk is truncated at some point, perhaps
by a binary companion, so we also assume the disk has a finite outer radius $R_{\rm d}$, which we set to $1000 R_\star$.

To complete our disk model, we have to specify the hydrogen model. We adopt a 25-level model of the H atom, where the first 12 levels are explicit NLTE levels, and the remaining 13 upper levels are implicit levels with LTE populations (with respect to the continuum). Each level has a consolidated statistical weight $g_n = 2n^2$. 

For the description of the central star we have to specify its effective temperature and surface gravity, which defines the Kurucz model atmosphere \citep{kur79,kur94} that we use as our input spectrum. We have to specify, in addition, the stellar radius and mass, which determine the critical velocity and scaleheight $H_0$.  Previously, we modeled the disk of $\zeta$ Tauri, B3\,IV \citep{woo97}, so we arbitrarily chose this spectral
type for our stellar parameters (see Table~\ref{tab1}).

\subsection{Grid Definition}

As described in section~\ref{MC}, the MC calculation is performed for a discretized model of the disk with each of the model unknowns (rates, temperature, level populations) being constant throughout a given cell. 
Thus, it is essential that the grid be constructed to minimize (as far
as practical) the variation of these unknowns within the cell.

The disk is divided into $n_{\rm r}$ radial shells, each of which are divided into $n_\mu$ cells in latitude and, if needed, into $n_\phi$ cells in azimuth.
We found a total of 40 cells in radius and 40 cells in $\mu$ to be a good compromise between usage of computer memory, speed of the code, and accuracy of the result.  The radial spacing is defined so that each cell has the same radial electron optical depth.  This approach has the drawback that the last cell becomes very large, so we further subdivide the last cell to properly sample the outer parts of the disk.  The $\mu$-spacing is defined in such a way that each cell represents an equal linear step in density, and, consequently, the cells are smaller close to the midplane and become larger for larger $z$. Also, the cells become larger with increasing radius because of the disk flaring. The disk is truncated in the $z$-direction when the density becomes $10^{-7}$ smaller than the value at the midplane. 

%-----------------------
\section{Results \label{results}}

In this section we explore the effects of the disk free parameters ($\rho_0$ and $\beta$) on the electron temperature and hydrogen level populations. We also investigate the effects of the temperature distribution on the IR excess. The parameters for all the models shown in the subsequent sections are listed in Table~\ref{tab2}.

\subsection{Ionization and Level Populations \label{LP}}

\begin{figure}[bp]
\plotone{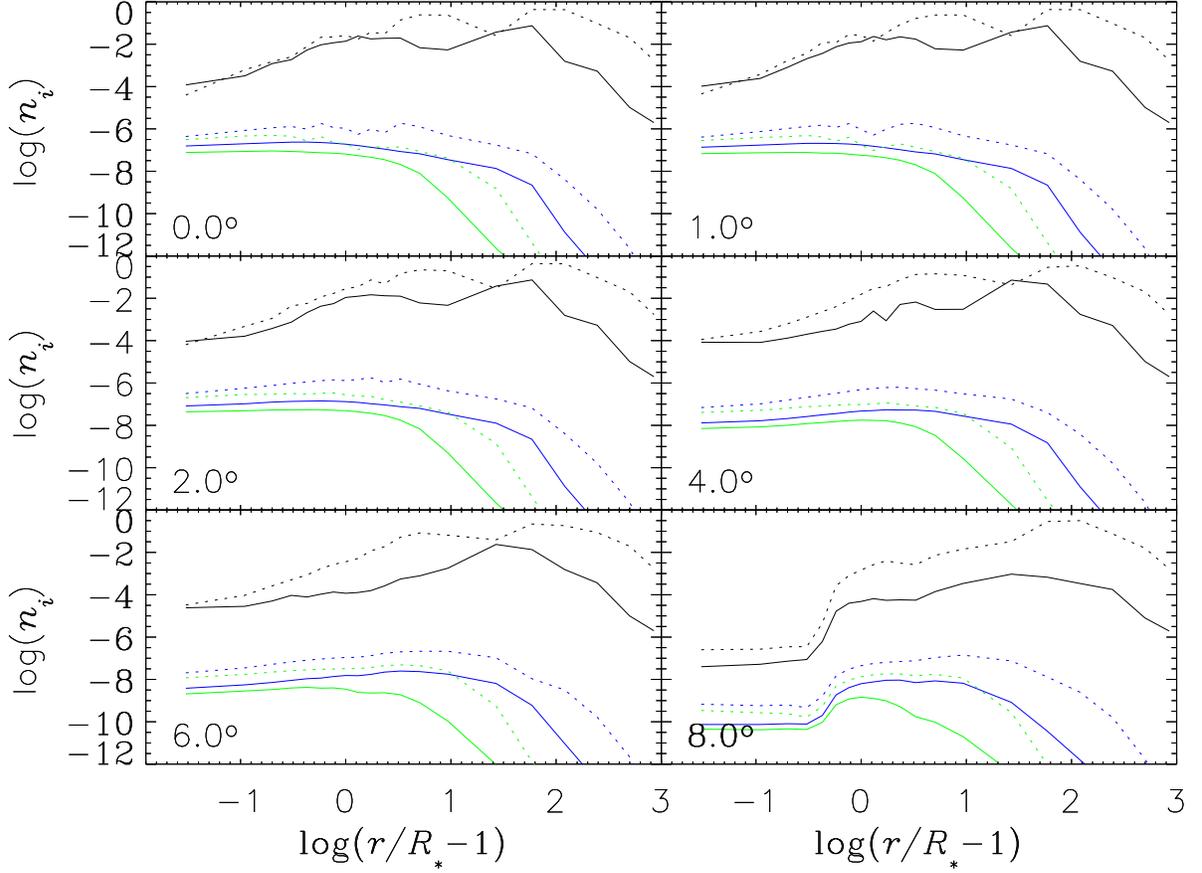}
\figcaption[]{Hydrogen Level Populations. Shown are the level populations for $n=1$ (black), $n=2$ (blue), and $n=3$ (green) as a function of radius. The solid curves are for model 01 (low density) while the dotted curves are for model 04 (high density).  Each panel shows the populations at a given latitude (angle above midplane), as indicated.
\label{LevelPopulations}}
\end{figure}

The $n=1$ and 2 level populations of hydrogen for models 01 and 04 (the density
extremes) are shown in Figure~\ref{LevelPopulations}, as a function of radius.
Each panel shows the level populations at different latitudes as indicated ($0^\circ$ corresponds to the disk midplane).  The approximations used in \citet{mil98,mil99a} make it difficult to perform a direct comparison between our results and theirs; however, they are qualitatively similar.

In general, the disk is almost completely ionized, with the exception of the midplane.  However, in none of the models does the midplane become completely neutral.  For model 04 (the highest density disk),
the neutral hydrogen fraction at the midplane has two maxima, one at $\varpi \approx 6 R_\star$ for which $n_1=20\%$ and the second at $\varpi \approx 50 R_\star$ where $n_1=44\%$. For model 01 (the lowest density model), the two maxima are also present, but closer to the star ($n_1=1\%$ at $\varpi \approx 3 R_\star$ and $n_1=6\%$ at $\varpi \approx 40 R_\star$).
These minima in the hydrogen ionization fraction indicate the location where the photoionizing mean intensity is smallest.  
The first minimum is also near the temperature minimum of the disk (see section~\ref{temperature}), so this position is likely to be the location of the low ionization state lines (like \ion{Fe}{2}) that form the shell features in Be spectra.

Another interesting feature of Figure~\ref{LevelPopulations} is the crossing of the low latitude $n=1$ curves.  At small radii, the lower density model (01) becomes more neutral than the higher density model (04). This apparent incongruity is a result of the higher density model being much hotter near
the star than the lower density model.

\subsection{Temperature \label{temperature}}

\begin{figure}[bp]
\plotone{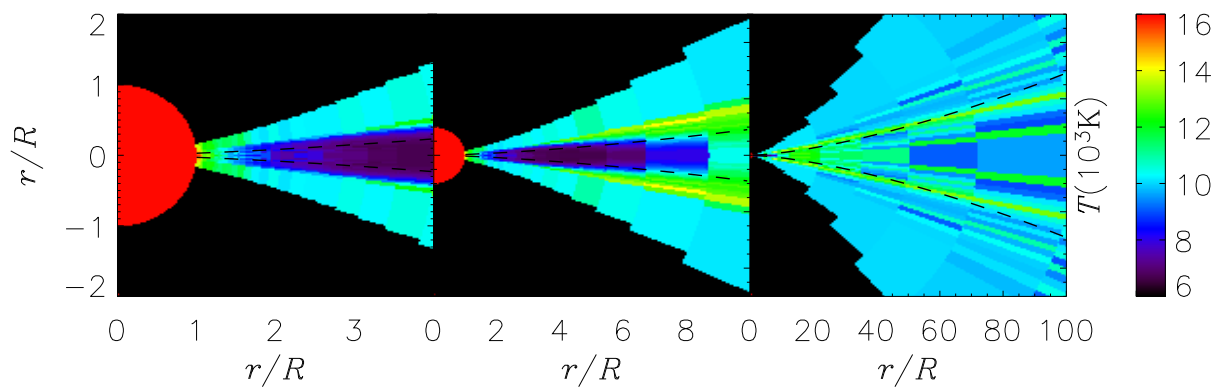}
\figcaption[]{Disk Temperature.  Shown is the temperature structure for model 02 at three radial scales. The dashed lines correspond to the curves 
$|z| = H(\varpi)$.
%For each panel the $y$ axis range from $-x_{\rm max}$ to $x_{\rm max}$.
\label{Temperature1}}
\end{figure}

The temperature structure of model 02 is shown in Figure~\ref{Temperature1}. 
The leftmost panel is the temperature in the vicinity of the star ($1<\varpi/R_\star<4$),
the center panel is the temperature for intermediate distances ($1<\varpi/R_\star<10$), and the rightmost panel is for large distances ($\varpi$ up to $100R_\star$).  In this figure, which generally is representative of all models, we observe two regions with very distinct temperature behaviors: the midplane region (between the dashed curves, defined by  $|z| \le H$), and the upper/lower layers of the disk ($|z| > H$).  In the upper layers, the temperature is nearly constant, averaging $10100 \pm 1300\;\rm K$ (about 50\% of $\teff$). 
%This result is consistent with the fact that  these regions are optically thin, both in the radial direction and in the $z$-direction. 
This region is optically thin for shorter wavelengths (typically $\lambda \lesssim 10 \um$), so it contributes very little to the disk emission at these wavelengths; however, it becomes optically thick for longer wavelengths.  Thus, this layer controls the long wavelength IR excess.

\begin{figure}[bp]
\epsscale{0.6}
\plotone{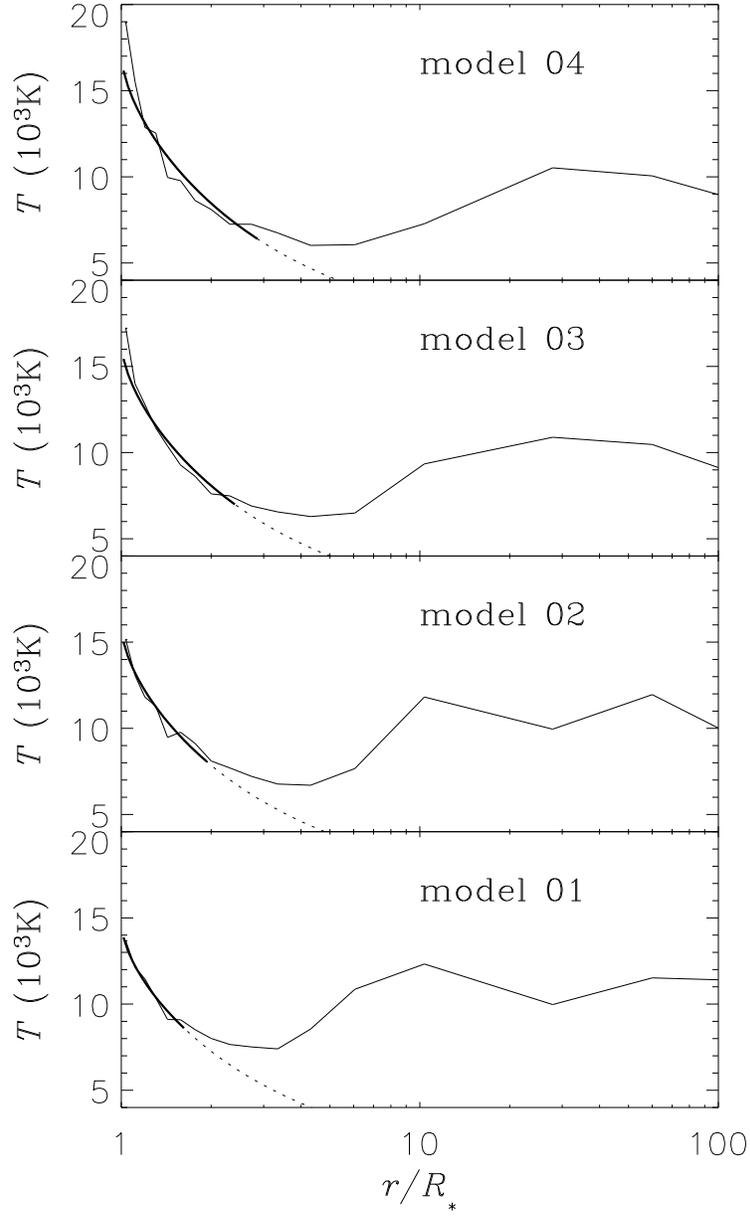}
\epsscale{1.0}
\figcaption[]{Mid-plane Temperature.  The radial dependence of the equatorial temperature is shown for models 01, 02, 03, and 04 (increasing density, respectively). The thick curve is a fit, for $\varpi$ between $R_\star$ and $\varpi_{\rm dep}$, to a flat blackbody reprocessing disk \citep{ada87}.
\label{Temperature2}}
\end{figure}

For the midplane region, the temperature structure is much more complex.
Beginning at the base of the disk, the temperature is high ($\approx15000\;\rm K$), quickly drops to a minimum temperature of $\approx 7000\;\rm K$ at $\varpi\approx4R_\star$, and rises back, reaching $\approx11000\;\rm K$ at $\varpi\approx10R_\star$, beyond which it becomes roughly isothermal.  
%These models differs only in the density scale. 
Other models show similar behavior, but the depth and location of the temperature minimum depends on the disk density (or equivalently, optical depth).  Figure~\ref{Temperature2} shows the mid-plane temperature as a function of increasing density (models 01--04).  Note that as the density
increases, the temperature minimum becomes deeper and occurs at larger radii.

Since the disk is {\it strongly non-isothermal} (unlike a stellar wind), it  is important to understand what determines its temperature structure.  In accretion disks around young stellar objects (YSOs), the temperature typically falls as a power law with radius, $T \propto r^{-m}$, with an
exponent $m$ in the range $1/2$--$3/4$, depending on the disk flaring.  This 
is because the surface of a very optically thick disk acts very much like a blackbody that reprocesses the incident starlight.  Since the star illuminates the surface at an increasingly oblique angle with radius,
the temperature falls faster than the usual $r^{-1/2}$ geometrical dilution
of the star.  For an infinitessimally thin, flat (i.e., not flared) blackbody disk, the temperature falls as
$r^{-3/4}$, but close to the star a detailed integration gives
\begin{equation}
T_{\rm d}(\varpi)  = 
%\left(
\frac{T_\star}{\pi^{1/4}}
%\right)^{1/4}
\left[
\sin^{-1}\left( \frac{R_\star}{\varpi} \right) - 
%\left(
\frac{R_\star}{\varpi}
%\right) 
\sqrt{1-\frac{R_\star^2}{\varpi^2}}
\right]
^{1/4},
\label{midplane_temperature}
\end{equation}
where $T_\star$ is the temperature of the radiation that illuminates the disk
\citep*[hereafter, ALS]{ada87}.
%This figure shows  an interesting result: although the qualitative behavior of the temperature is similar for the different models, the actual values are very different, the higher the density the larger the temperature at the base of the star.

Using this function, we fit the mid-plane temperatures (close to the star) for each model. The resulting curves for models 01--04 are shown in Figure~\ref{Temperature2} (thick lines), while the values of $T_\star$ along with $\varpi_{\rm dep}$ (the outer radius of the fitting interval) are listed in Table~\ref{tab2}.   We see that eq.~(\ref{midplane_temperature}) reproduces the initial temperature decrease quite well.  However, $T_\star$ is \emph{larger than $\teff$} for all models (except the lowest density model 01), and the larger the density, the larger the corresponding $T_\star$.  It is evident, 
therefore, that this behavior is due to back-warming of the star (photons that hit the star are reemitted locally to simulate this effect).
This result reveals a small inconsistency in our model: the back-warming is sufficiently large that the stellar photospheric temperature should be increased near the equator.  We did not attempt to fully model this effect; 
we only reemitted the photons without raising the temperature.

Another interesting feature of Figure~\ref{Temperature2} is that, as the density increases, the departure point, $\varpi_{\rm dep}$ (the location where the MC temperature curve departs from eq.~[\ref{midplane_temperature}]) shifts to larger radii. This suggests that $\varpi_{\rm dep}$ is controlled by the disk optical depth. Indeed, we found that $\varpi_{\rm dep}$ correlates very well with the vertical electron scattering optical depth 
\begin{equation}
\tau_{\rm e}(\varpi_{\rm dep}) = \int_{-\infty}^\infty  \sigma_{\rm T} 
\frac{\gamma}{\mu m_{\rm H}} \rho(\varpi_{\rm dep},z)  dz,
\end{equation}
where $\sigma_{\rm T}$ is the electron scattering cross section, $\gamma$ is the number of electrons per ion, $\mu$ is the mean molecular weight, and $m_{\rm H}$ the mass of the hydrogen atom. The values of $\tau_{\rm e}(\varpi_{\rm dep})$ for each model are listed in Table~\ref{tab2}.  We conclude that the temperature stops falling because the disk becomes (vertically) optically thin.
Once the disk becomes optically thin, ionizing photons from high stellar latitudes begin to penetrate the disk, heating it back up to the optically thin radiative equilibrium temperature at large radii. Beyond this location the disk becomes roughly isothermal, behaving much like a optically thin stellar wind.

The behavior of the midplane temperature can be summarized as follows: close to the star, the temperature is described by a flat blackbody reprocessing disk, eq.~(\ref{midplane_temperature}), with $T_\star \approx \teff$ for low $\rho_0$ and $T_\star > \teff$ for larger $\rho_0$. When the vertical electron optical depth drops below $\tau_{\rm e} \approx 0.075$, the temperature curve departs from eq.~(\ref{midplane_temperature}) and quickly returns to the optically thin radiative equilibrium temperature (about 60\% $\teff$) for $\varpi \gtrsim 10$--$30R_\star$.

We end this section with a comparison between our results and those of \citet{mil99a,mil99b}.  For the optically thin parts of the disk, our results agree --- we also find a nearly constant temperature of about 60\% $\teff$.  For the midplane, however, our results differ. Their temperature in the outer parts of the disk is in general lower than ours, while in the inner parts, their results are often higher. In addition, their mid-plane temperature profiles initially drop much more quickly than ours, and they sometimes have a local maximum at intermediate radii that is not seen in our results.  
Nonetheless, we do agree qualitatively that there is temperature minimum at small radii in the disk mid-plane.

\subsection{Disk Flaring}
Let us now investigate the effects of the disk flaring parameter $\beta$ on the temperature structure.  Experience with the flared disks of YSOs shows that the flaring controls the amount of heating versus radius, which determines the temperature versus radius \citep[e.g.,][]{ken87}.  In general, one finds that the smaller the flaring, the steeper the temperature gradient.

\begin{figure}[bp]
\plotone{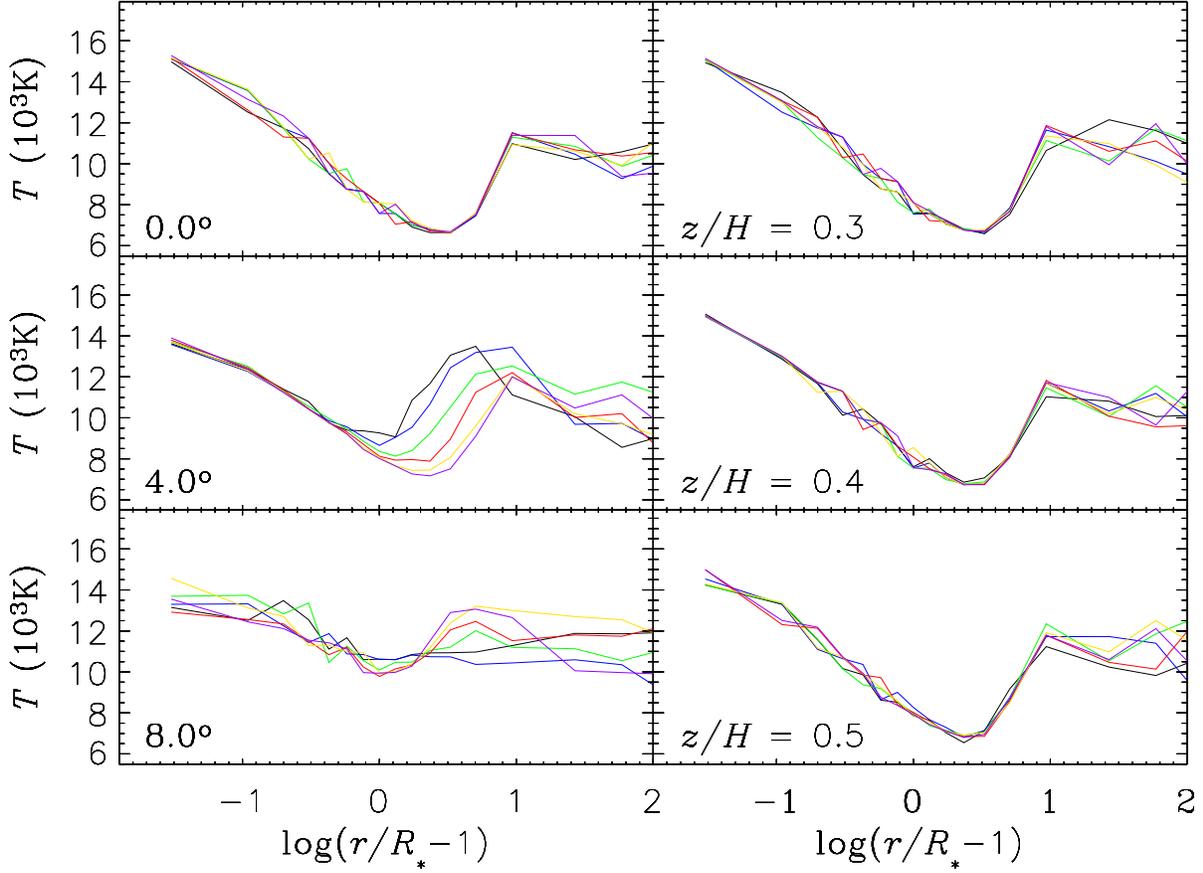}
\figcaption[]{
Effects of Disk Flaring.  The temperature versus
radius is shown for six values of the disk flaring exponent, $\beta=1.0$--1.5 in steps of 0.1 (black, blue, green, red, orange, and purple lines, respectively). The different panels show the temperature along rays with a given latitudelatitude (left) and for a given fraction of the disk scaleheight
(right).
\label{Temperature3}}
\end{figure}

We do not, however, observe this in our results.  Figure~\ref{Temperature3} shows the temperature as a function of radius for models 05, 06, 07, 08, 09 and 02, which have $\beta$ ranging from 1.0 (wedge-shaped disk) to 1.5 (radially isothermal disk).  The left side shows the results for three different 
latitudes.  We see that the mid-plane temperature is very similar for all the models, but for the other latitudes, the temperatures are quite different. This difference, however, is a result of the different 
scaleheights probed by each latitude ray.  If instead, we plot the temperatures at fixed fractions of the scaleheight above the mid-plane
(right side), the curves are all very similar.  We conclude that
the disk flaring does not influence the disk heating versus radius.

There are probably two reasons for this: 1) at small radii, the disk surface
is much less obliquely illuminated, and the direction perpendicular to the 
disk surface is insensitive to $\beta$ (this is why the ALS model works well 
for this region); and 2) the outer parts of the disk are optically thin, so
they heat up to the optically thin radiative equilibrium temperature,
regardless of the disk flaring.

\subsection{IR excess \label{IR}}

\begin{figure}[bp]
\plotone{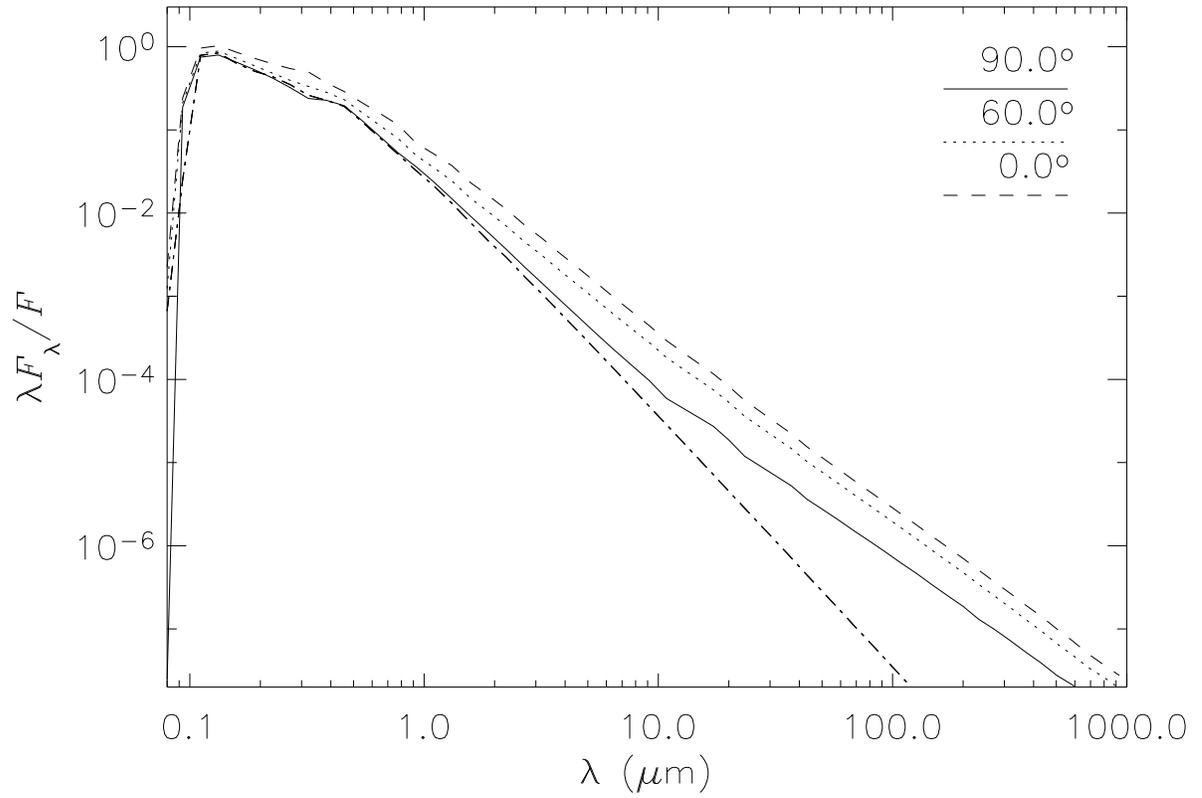}
\figcaption[]{Spectral Energy Distribution.  Shown is the emergent flux for Model 02, viewed at three different inclinations, as indicated. The dash-dotted line indicates the input flux from the star (a 19000\,K Kurucz model 
atmosphere).
\label{SED}}
\end{figure}

The spectral energy distribution for Model 02 in the 0.08--1000$\;\mu\rm m$ wavelength range is shown in Figure~\ref{SED}.  Interestingly, in addition to 
the usual IR excess, there is a small 
UV excess (in the Balmer continuum) for pole-on viewing angles.  Just longward
of the Lyman jump, this UV excess is a result of electron scattering in the 
disk, which redirects some of the stellar flux into the polar direction.  As 
the wavelength increases, bound-free emission begins to dominate, so the 
pole-on excess increases until the Balmer jump occurs.  Beyond the Balmer 
jump, the f-f emission (and opacity) becomes increasingly important.  
As the envelope becomes optically thick, the disk develops a 
pseudo-photosphere that grows in size with increasing wavelength, producing 
the large IR excess.  Since the disk is geometrically thin, the area of the 
pseudo-photosphere is smaller viewed edge-on than pole-on.  As a result, the
edge-on IR flux is about 10--50\% of the pole-on flux, depending on wavelength.
Note that the slope of the IR excess also depends on inclination angle.

As mentioned in the introduction, the slope of the IR excess has been used 
by \citet{wat86} to observationally measure the radial density exponent for 
the circumstellar envelopes of Be stars.  However, as demonstrated in
section~\ref{temperature} \citep[also see][]{mil98}, the temperature varies
throughout the disk by more than a factor of two, which raises questions about the validity and accuracy of Waters' results.  

\begin{figure}[bp]
\epsscale{0.5}
\plotone{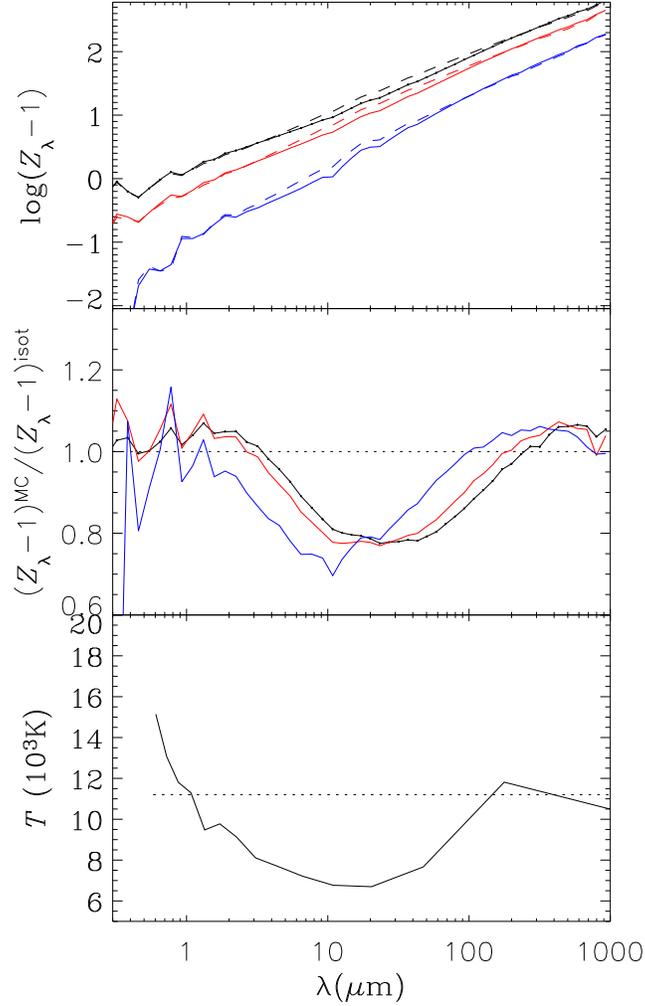}
\epsscale{1.0}
\figcaption[]{IR Excess for Model 02. The top panel compares the radiative 
equilibrium IR excess (solid lines) to that of an equivalent isothermal model
(dashed lines) for three inclinations: pole-on (black), $60^\circ$ (red), and 
edge-on (blue).  The isothermal model has a temperature $T_{\rm iso}=0.6\teff$, which yields the best fit.  The middle panel shows the ratio of the IR excesses for the radiative equilibrium and isothermal models, while the bottom panel
shows the disk mid-plane temperature at $R_{\rm eff}(\lambda)$, the radius
of the pseudo-photosphere as a function of wavelength (see text).  Note that 
the deviation between the radiative equilibrium and isothermal excesses roughly
tracks the shape of the mid-plane temperature.
\label{IR1}}
\end{figure}

To isolate and assess the effects of the temperature structure, we compare the IR excess,
\begin{equation}
Z_\lambda-1 \equiv (F_\lambda-F_\lambda^\star)/F_\lambda^\star,
\end{equation}
of the radiative equilibrium model to that of an equivalent isothermal model (i.e., one with the same density and velocity).  For model 02, the best-fitting isothermal equivalent has a kinetic temperature of 
$11200\rm K$ (about 60\% of $\teff$), and its comparison to the radiative equilibrium
model is shown in Figure~\ref{IR1}.  The middle panel, which shows the ratio between the radiative equilibrium and isothermal IR excesses, indicates how 
well the isothermal model reproduces the full radiative equilibrium solution. Surprisingly, there is little deviation ($<20\%$) for all inclination angles. In detail, the excesses are nearly identical for both short and long
wavelengths, while the largest differences occur for intermediate wavelengths, $1 \lesssim \lambda \lesssim 100\;\mu \rm m$.

The reason an isothermal model works so well at very long wavelengths 
($\lambda \gtrsim 100\;\mu \rm m$) is because the disk is so optically thick
that its pseudo-photosphere occurs in the isothermal region.  For the
pole-on case, the effective radius, $R_{\rm eff}$, of the pseudo-photosphere 
is the location where the vertical optical depth $\tau_\lambda(R_{\rm eff}) = 1$.  It is straightforward to show (see Appendix~\ref{App_IR}, eq.~[\ref{app_reff}]) 
that  
\begin{equation}
R_{\rm eff}(\lambda) \propto \lambda^{\frac{u+2}{2n-\beta-3s/2}},
\end{equation}
where $s$ and $u$ are the exponents of the power-laws describing the 
temperature and Gaunt factors, respectively (see eqs.~[\ref{tpowerlaw}] and 
[\ref{gpowerlaw}]). 
%For the density of model 02, $R_0 = 1.268 R_\star$.
Using $\beta=1.5$, $n=3.5$, $s=0$, and $u=0.23$, we find (for model 02) that
\begin{equation}
R_{\rm eff}(\lambda) = 1.27 R_\star (\lambda/1\mu{\rm m})^{0.41}.
\label{R_vs_L}
\end{equation}
For $\lambda = 1000 \;\mu \rm m$, this gives $R_{\rm eff} \approx 21 R_\star$,
verifying that the effective photospheric radius is indeed in the isothermal region for large wavelengths.

At shorter wavelengths ($\lambda \lesssim 100 \;\mu \rm m$), the effective 
radius is in the non-isothermal region.  As a result, the isothermal IR excess 
begins to differ from the radiative equilibrium excess.  These differences directly reflect the temperature structure of the disk midplane.  Using eq.~(\ref{R_vs_L}) to translate from $R_{\rm eff}$ to wavelength, we plot 
(in the bottom panel of Fig.~\ref{IR1}) the midplane temperature 
$T(R_{\rm eff})$ as a function of wavelength.  Note the similarity between
the flux departures (middle panel) and the midplane temperatures (bottom 
panel).  In particular, the largest difference occurs near the temperature
minimum of the disk.  This suggests the interesting, albeit technically
challenging, possibility of inverting the observed IR excess to obtain the midplane temperature profile.  The flux excess in the optically thick
limit (c.f., eq.~[\ref{Zthick}]), is given approximately by
\begin{eqnarray}
Z_\lambda-1 &\approx& \pi B_\lambda(R_{\rm eff}) R^2_{\rm eff} / 
                    F^\star_\lambda\ , \nonumber \\
            &\approx& \frac{T(R_{\rm eff})}{\teff} 
                   \left(\frac{R_{\rm eff}}{R_\star}\right)^2\ .
\end{eqnarray}
Consequently, the excess ratio
\begin{equation}
\frac{(Z_\lambda-1)^{\rm MC}}{(Z_\lambda-1)^{\rm iso}} \sim
\frac{T(R^{\rm MC}_{\rm eff})}{T_{\rm iso}} 
                   \left(\frac{R^{\rm MC}_{\rm eff}}{R^{\rm iso}_{\rm eff}}\right)^2\ ,
\end{equation}
which demonstrates why the flux departures track the the midplane
temperatures.  Although it appears that the flux departure depends linearly
on the midplane temperature, one must recall that $R_{\rm eff}$ is 
temperature-dependent via the opacity.  As the midplane radiative 
equilibrium temperature decreases, the opacity increases (see 
eq.~[\ref{app_kappa}]), making $R^{\rm MC}_{\rm eff}$ larger than the 
corresponding isothermal value.  This increase in the effective area of the 
disk pseudo-photosphere partially offsets the decrease 
of disk source function, $B_\lambda(R_{\rm eff})$.  The net result is
that the flux departure is smaller than naively expected.  Thus detailed 
modeling will be required to invert the flux departures to obtain the
disk midplane temperature versus radius.

\begin{figure}[bp]
\plotone{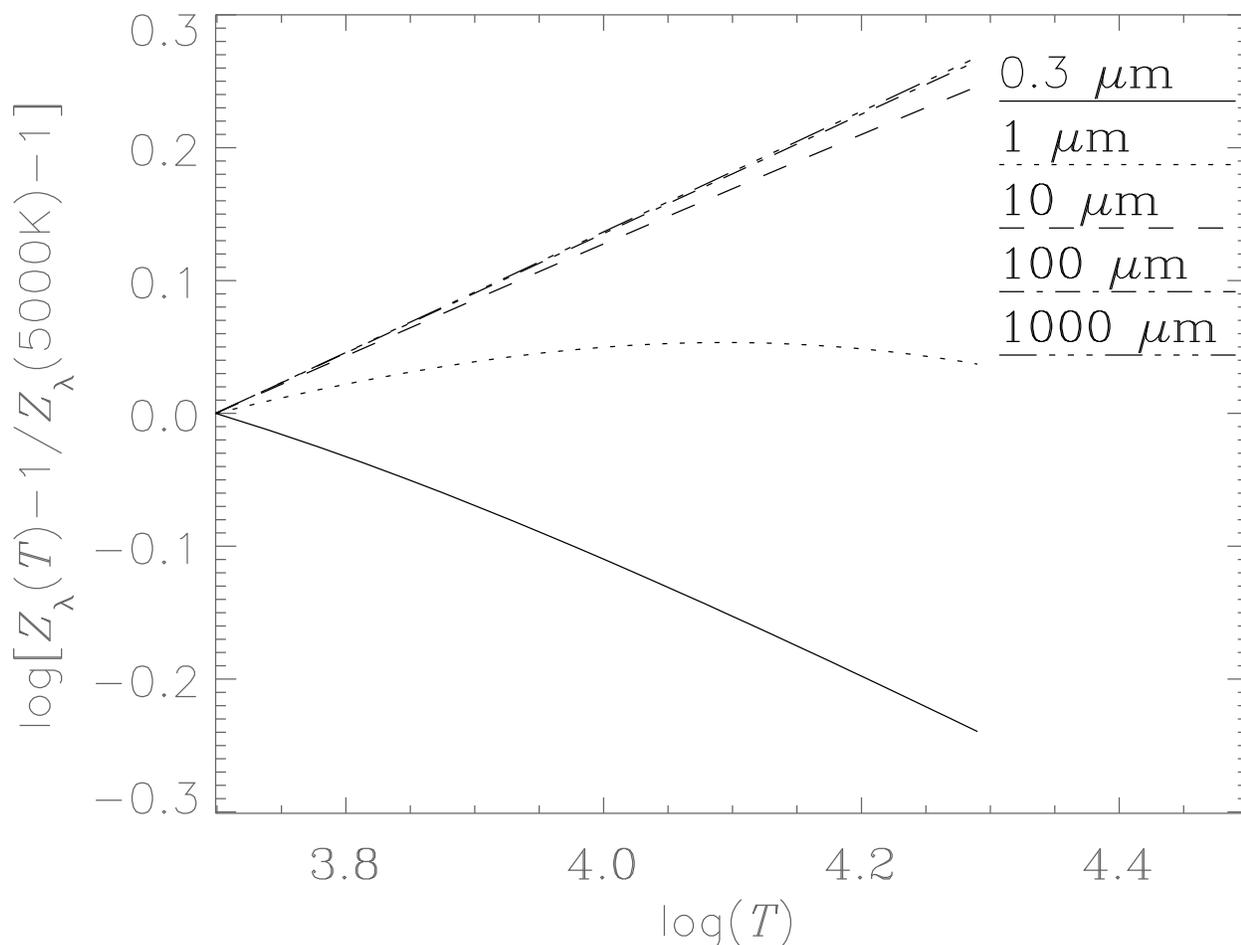}
\figcaption[]{Temperature-dependence of the IR Excess.  The IR excess as
a function of temperature is shown for different wavelengths, as indicated. Each curve was normalized to the excess at $5000\;\rm K$.  Note that, for the optically thin (shorter) wavelengths, the excess decreases with increasing
temperature, while for the optically thick (longer) wavelengths, the excess increases.
\label{E_vs_T}}
\end{figure}

The above discussion explains the flux differences in the 1--$100\;\mu \rm m$ 
range.  It does not, however, explain why the isothermal and radiative equilibrium models agree for wavelengths $\lambda \lesssim 1\;\mu \rm m$, 
where the midplane temperature continues to rise well above the isothermal
model (see bottom panel of Fig.~\ref{IR1}).  To help understand this result,
we show the wavelength-behavior of the IR excess versus temperature in
Figure~\ref{E_vs_T}.  As discussed previously, the IR excess increases
with temperature for optically thick wavelengths.  However for optically
thin wavelengths, the excess decreases owing to the smaller f-f 
emissivity at lower temperatures (see eq.~[\ref{Zthin}]).  As a consequence, 
marginally thick wavelengths have little temperature dependence, as shown 
by the $1\;\mu{\rm m}$ curve (dotted line) in Figure~\ref{E_vs_T}.  Since
$(Z_\lambda-1)$ is a weak function of $T$ in the 0.5--$1\;\mu \rm m$ 
range, isothermal models reproduce the flux excess quite well in this
region.

\begin{figure}[bp]
%\epsscale{0.5}
\plotone{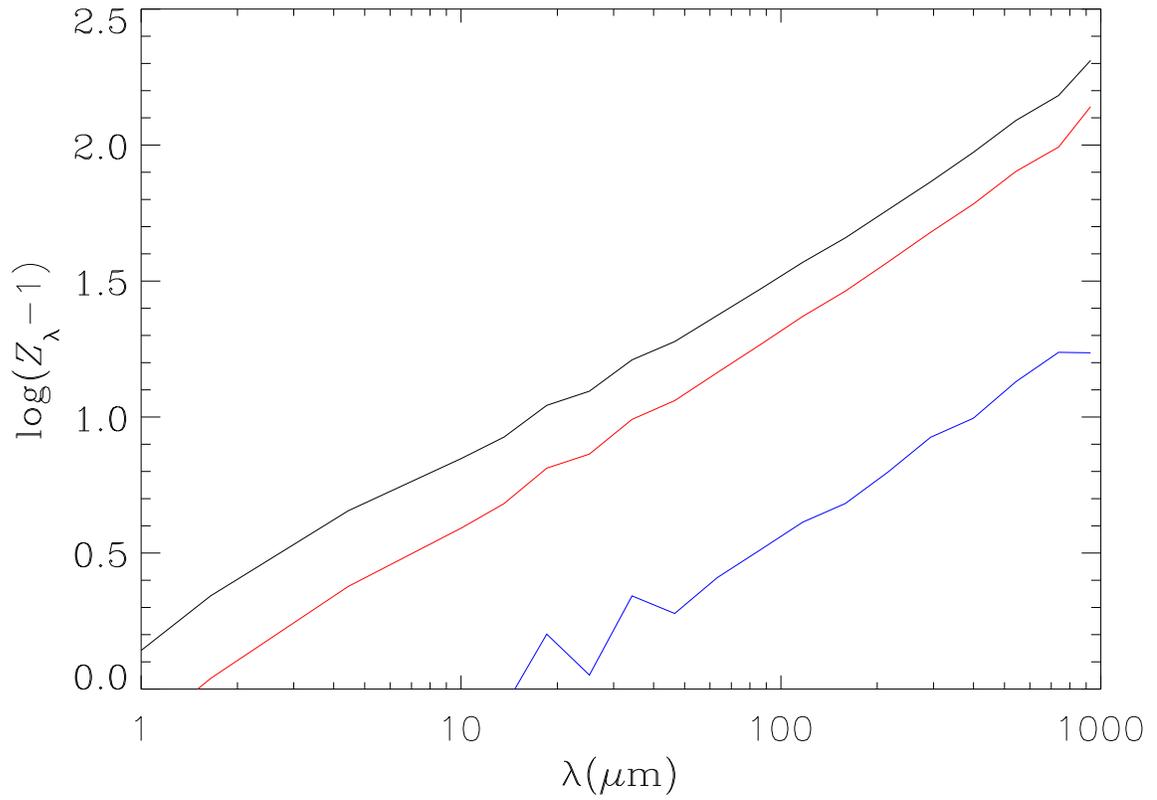}
\epsscale{1.0}
\figcaption[]{Infrared excess vs. wavelength for model 05 ($\beta = 1.0$). Shown are results for three inclinations: pole-on (black), $60^\circ$ (red), and 
edge-on (blue).  
\label{SED2}}
\end{figure}

In general, we have seen that radiative equilibrium models are 
well-approximated by an equivalent isothermal model with a temperature 
$T_{\rm iso} = 0.6 \teff$.  However, the temperature structure is not the
only factor influencing the IR SED.  As shown in the appendix, the slope
of the excess depends on the disk flaring parameter $\beta$ (see eq.~[\ref{Zthick}]).  Furthermore, as may be seen in Figures~\ref{SED} and \ref{IR1}, 
the slope of the IR excess changes with inclination angle (and wavelength).
The reason for the inclination-dependence is that, as $R_{\rm eff}$ changes with wavelength, the shape of the flared disk pseudo-photosphere is 
not self-similar (except for the pole-on case).  In contrast for an unflared disk ($\beta=1$), the shape is self-similar (with wavelength) for
all inclination angles, so the slope of the IR excess is independent of inclination, as demonstrated in Figure~\ref{SED2}.  We conclude that the primary limitation of Waters' method for determining the disk density distribution from the slope of the IR excess is not the isothermal assumption; 
it is the assumption of a particular geometry (e.g., a constant opening angle) for the disk.

\section{Discussion and Conclusion \label{discussion}}

In this paper, we present a new 3-D NTLE Monte Carlo code that fully solves 
the radiation transfer and radiative equilibrium for arbitrary gas density and 
velocity distributions.  Even though our method was developed independently, it is similar in many respects to the Monte Carlo transition probability method of \citet{luc02,luc03}.  However, our method differs from Lucy's in a 
fundamental way, which is that our photon absorption and reemission mechanisms
are uncorrelated --- we sample the entire emissivity distribution regardless
of the absorption process.  This allows us to dispense with Lucy's macro atoms, along with their associated internal transitions and Monte Carlo transition probabilities.
%, making our method easier to implement.  
Tests of our code against an independent (1-D spherically symmetric) NLTE code show that our 
method rapidly converges to the correct results, typically requiring only
a few iterations.  This rapid convergence also occurs for the non-trivial
2-D disk geometry investigated here.

Since the primary purpose of our NTLE Monte Carlo method is the study of
gaseous circumstellar disks, we performed an initial investigation of the 
temperature and ionization structure of Be star disks. Our results for the disk temperature show that its behavior is a hybrid between that of optically 
thick disks of Young Stellar Objects and optically thin winds from Hot Stars.  
We find that the temperature initially drops very quickly close to the stellar 
photosphere, and the temperature profile is well-described by a flat blackbody
reprocessing disk \citep{ada87}.  When the disk becomes optically thin 
vertically, the temperature departs from this curve and rises back to the 
optically thin radiative equilibrium temperature, which is approximately
constant in the winds of Hot Stars.  We find that the average temperature 
in these isothermal regions is about 60\% of $\teff$, which is cooler than the 
80\% value frequently quoted in the literature  \citep[e.g.,][]{wat86}.
Similarly, the upper layers of the disk heat up to the optically thin radiative equilibrium temperature, which again is approximately isothermal.  This 
behavior is analogous to the hot upper layers of YSO disks 
\citep[see][]{chi97}.  However, unlike YSO disks whose flaring controls their heating, we find that the disk temperature of Be star disks is influenced very little by the disk flaring.

As to be expected, we find that the disk is entirely ionized, but there is a region close to the midplane (from about 3--6$\;R_\star$ outwards) for which the neutral hydrogen fraction ranges from a few~\% up to 50\%, depending on the model density.  This site of low ionization coincides with the temperature minimum in the disk midplane.  Because the location of the temperature minimum is controlled by the vertical optical depth, we find that as the density increases, the temperature minimum occurs at larger radii, and the disk becomes more neutral owing to
the decrease in photoionization at high optical depths.  This suggests the temperature minimum also is likely to be the site for the formation of low ionization state lines (such as \ion{Fe}{2}) that produce the shell features observed in Be stars.

Finally, to investigate the consequences of the non-isothermal temperature 
structure on the IR excess of Be stars, we calculated the SEDs for our models 
and compared them to equivalent isothermal models. We find the effects of the temperature structure are small ($< 20\%$ change in the flux), and that the IR 
excess is well-described by an isothermal model whose temperature is about 
$0.6 \teff$.  Hence the slope of the IR excess is less sensitive to the
disk temperature structure than it is to the disk geometry (e.g., inclination 
angle and flaring).  We conclude the main limitation for using the IR excess 
to determine disk properties is the use of an assumed disk geometry.

In the next paper of this series, we will address the question of disk
geometry by self-consistently determining both the disk temperature and 
density.  To do so, we will use the radiative equilibrium temperatures
to solve the vertical hydrostatic equilibrium structure for a Keplerian disk with steady-state viscous accretion (or decretion).  This self-consistent
solution for the disk structure and radiative transfer will permit us to
quantitatively test whether or not the SEDs and intrinsic polarization
of Be stars can be produced by a viscous decretion disk.

\acknowledgments{
The authors are grateful to the referee, Dr. L.B. Lucy, for his useful comments.
The authors acknowledge support from NSF grants AST-9819928
and AST-0307686. 
A.C.C. acknowledges partial support from Fapesp 04/07707-3.}

\appendix 

\section{IR Excess of Pole-on Flared Disks\label{App_IR}}

Following a method similar to that employed by \citet{wat86}, we consider a 
star surrounded by a fully ionized disk with density given by 
eqs.~(\ref{verticalrho}) and (\ref{scaleheight}). 
Let us assume a power-law distribution for the temperature
\begin{equation}
%T(\varpi)=T_0\left(\frac{\varpi}{R_\star}\right)^{-s}\ ,
T(\bw)=T_0\bw^{-s}\ ,
\label{tpowerlaw}
\end{equation}
where $\bw=\varpi/R_\star$, as well as a power-law for the free-free plus bound-free Gaunt factors
\begin{equation}
g_\lambda=g_0\left(\frac{\lambda}{\lambda_0}\right)^{u}.
\label{gpowerlaw}
\end{equation}

Since we assume the disk is viewed pole-on, the intensity for $\bw \le 1$ is not changed by the presence of the disk. Hence, the intensity $I_\lambda$ may be written as
\begin{equation}
I_\lambda(\bw) =
\left\{
\begin{array}{ll}
B_\lambda^\star(\teff)
 & 
 \mbox{($\bw \le 1$),} \\
B_\lambda^{\rm d}(T)\left[1- e^{-\tau_\lambda(\bw)} \right]
 &
\mbox{($\bw > 1$),}
\end{array}
\right.
\end{equation}
where we assumed, for simplicity, that the disk is vertically isothermal and
that the star emits a blackbody spectrum (for the spectral range of interest, $\lambda > 1 \um$, a hot star emits a spectrum that is similar to a blackbody).

The optical depth along the line of sight is
\begin{equation}
\tau_\lambda(\bw) =
\int_{-\infty}^{\infty}\kappa_\lambda(\bw) dz, \label{app_tau_def}
\end{equation}
where the opacity
\begin{equation}
\kappa_\lambda = 3.69\;10^8 T^{-1/2} c^{-3} \lambda^{3}
\left( 1-e^{-hc/\lambda kT} \right) g_\lambda \frac{\gamma^2 \rho^2}{\mu^2 m^2_{\rm H}}.
\label{app_kappa}
\end{equation}
For long wavelengths and the typical temperature range of Be star disks (several thousand K), the opacity can be written as
\begin{equation}
\kappa_\lambda = \kappa_0
%\left(\frac{\varpi}{R_\star}\right)^
\bw^
{3s/2-2n}e^{-z^2/H^2}\lambda^{2+u},
\end{equation}
with 
\begin{equation}
\kappa_0 = 1.77\;10^{-2} T_0^{-3/2} c^{-3} g_0  \lambda_0^{-u}  \frac{\gamma^2 \rho_0^2}{\mu^2 m^2_{\rm H}}.
\end{equation}
Performing the integration in eq.~(\ref{app_tau_def}), we find
\begin{equation}
\tau_\lambda(\bw) = \tau_\lambda^\prime \bw^{3s/2-2n+\beta},  \label{app_tau}
\end{equation}
with $\tau_\lambda^\prime \equiv \sqrt{\pi} \kappa_0 H_0 \lambda^{2+u}$.

The total flux (star plus disk) is given by
\begin{equation}
F_\lambda = 
\int_0^{2\pi} d\phi
\int_0^{\bar{R}_{\rm d}} I_\lambda(\bw) \bw d\bw. \label{app_flux}
\end{equation}
Defining the \emph{excess flux} $Z_\lambda - 1\equiv (F_\lambda-F_\lambda^\star)/F_\lambda^\star$, we have
\begin{equation}
Z_\lambda-1 = 
%\frac{2}{R_\star^2} \int_{R_\star}^{R_{\rm d}} 
2 \int_{1}^{\bar{R}_{\rm d}} 
\frac{B_\lambda^{\rm d}}{B_\lambda^\star} 
\left(1- e^{-\tau_\lambda^\prime \bw^{3s/2-2n+\beta}} \right) \bw\,d\bw\ , 
\label{app_excess1}
\end{equation}
which can be written in the Rayleigh-Jeans limit as
\begin{equation}
Z_\lambda-1 =  
%\frac{2}{R_\star} 
2
\frac{T_0}{\teff} 
\int_{1}^{\bar{R}_{\rm d}} 
%\left(\frac{\varpi}{R_\star}\right)^{1-s} 
\bw^{1-s}
\left(1- e^{-\tau_\lambda^\prime 
%\left(\varpi/R_\star\right)
\bw
^{3s/2+\beta-2n}} \right) d\bw.
\label{app_excess2}
\end{equation}

Equation~(\ref{app_excess2}) can be integrated numerically to obtain the flux excess. However in the limiting cases of optically thin and optically thick disks, useful analytical approximations can be obtained.
In the optically thin case, 
$1- \exp\left(-\tau_\lambda^\prime  \bw^{3s/2-2n+\beta}\right) 
\approx \tau_\lambda^\prime  \bw^{3s/2-2n+\beta}$, so eq.~(\ref{app_excess2}) becomes
\begin{equation}
Z_\lambda-1 = 2 \frac{T_0}{\teff}
\frac{\bar{R}_{\rm d}^{2+s/2+\beta-2n}-1}{2+s/2+\beta-2n}
\tau_\lambda^\prime.
\end{equation}
Because $\tau_\lambda^\prime \propto T_0^{-3/2}$, it follows that the IR excess in the optically thin case will scale with $T_0$ as
\begin{equation}
Z_\lambda-1 \propto T_0^{-1/2}.
\label{Zthin}
\end{equation}

In the optically thick limit, a useful approximation is to consider that most of the radiation comes from a pseudo-photosphere for which the effective radius 
$\bar{R}_{\rm eff}(\lambda)$ is defined so that
\begin{equation}
\tau_\lambda(\bar{R}_{\rm eff}) \equiv 1. 
\end{equation}
Using eq.~(\ref{app_tau}), we determine that
\begin{equation}
\bar{R}_{\rm eff}(\lambda) = \bar{R}_0 \left( \frac{\lambda}{c}\right)^{\frac{u+2}{2n-\beta-3s/2}}\ ,
\label{app_reff}
\end{equation}
where 
\begin{equation}
\bar{R}_0 = \frac{(\sqrt{\pi}\kappa_0 H_0)^{(2n-\beta-3s/2)^{-1}}}{R_\star}. 
\end{equation}
With 
$\bar{R}_{\rm eff}$ so defined, the flux excess for the optically thick disk 
is 
\begin{eqnarray}
Z_\lambda-1 &=& 2  \frac{T_0}{\teff} \int_{1}^{\bar{R}_{\rm eff}} 
%\left(\frac{\varpi}{R_\star}\right)
\bw
^{1-s} d\bw, \nonumber \\
            &=& \frac{2}{2-s} \frac{T_0}{\teff} 
\left[
\bar{R}_0^{2-s} \left(\frac{\lambda}{c}\right)^{\frac{(u+2)(2-s)}{2n-\beta-3s/2}} - 1
\right].
\label{Zthick}
\end{eqnarray}

\clearpage

\clearpage

\clearpage

\begin{deluxetable}{ll}
\tablecaption{Adopted Stellar Parameters. \label{tab1}}
\tablewidth{0pt}
\tablehead{\colhead{Parameter} & \colhead{Value}
}
\startdata
Spectral Type \ldots	&	B3~IV \\
$R_\star$ \dotfill		&	$5.6 R_\sun$ \\
$\teff$ \dotfill		&	19000 K \\
$\log g$ \dotfill		&	4 \\
$M_\star$ \dotfill		&	$9.2 M_\sun$\\
$v_{\rm crit}$ \dotfill	&	$560 \mathrm{\;km\;s^{-1}}$
\enddata
\end{deluxetable}

\begin{deluxetable}{clcccc}
\tablecaption{Disk Model Parameters. \label{tab2}}
\tablewidth{0pt}
\tablehead{
\colhead{Model} &
\colhead{$\rho_0$} &
\colhead{$\beta$} &
%\colhead{$\langle T \rangle_{|z| < H(\varpi)}^{d<10}$} &
%\colhead{$\langle T \rangle_{|z| < H(\varpi)}$} &
%\colhead{$\langle T \rangle_{|z| > H(\varpi)}$} &
\colhead{$T_\star$} &
\colhead{$\varpi_{\rm dep}$} &
\colhead{$\tau_{\rm e}(\varpi_{\rm dep})$}
\\
\colhead{} & \colhead{$(\rm g\;cm^{-3})$} & \colhead{} &
\colhead{(K)} & \colhead{($R_\star$)}   & \colhead{}
}
\startdata
01	&	$1.66 \; 10^{-11}$ 	&	1.5	& 
%$12100 \pm 2600$ &  $11800 \pm 2600$ &  $11000 \pm 1800$  &  
17600 & 1.7  &   .063 \\
02	&	$4.15 \; 10^{-11}$ 	&	1.5	&	
%$9500 \pm 4700$ &  $12100 \pm 2300$ &  $11200 \pm 1700$  & 
19100 &  2.0  &   .081 \\
03	&	$8.39 \; 10^{-11}$ 	&	1.5	&	
%$7500 \pm 6400$ &  $12200 \pm 2200$ &  $11600 \pm 1600$  & 
19600 &  2.5  &   .072 \\
04	&	$1.66 \; 10^{-10}$ 	&	1.5	&	
%$6600 \pm 7300$ &  $12600 \pm 2100$ &  $11800 \pm 1600$  & 
20500 &  3.0  &   .081 \\
05	&	$4.15 \; 10^{-11}$ 	&	1.0	&
%$XX \pm XX$ &  $XX \pm XX$ &  $XX \pm XX$  &  
18600 &   2.0  &   .083 \\
06	&	$4.15 \; 10^{-11}$ 	&	1.1	&
%$XX \pm XX$ &  $XX \pm XX$ &  $XX \pm XX$  &  
18800 &   2.0  &   .078 \\
07	&	$4.15 \; 10^{-11}$ 	&	1.2	&
%$XX \pm XX$ &  $XX \pm XX$ &  $XX \pm XX$  &  
18200 &   2.0  &   .077 \\
08	&	$4.15 \; 10^{-11}$ 	&	1.3	&
%$XX \pm XX$ &  $XX \pm XX$ &  $XX \pm XX$  &  
18700 &   2.0  &   .080 \\
09	&	$4.15 \; 10^{-11}$ 	&	1.4	&
%$XX \pm XX$ &  $XX \pm XX$ &  $XX \pm XX$  &  
18600 &   2.0  &   .085 \\
\enddata
\end{deluxetable}

\end{document}